学号 2014301000121

密级

# 武汉大学本科毕业论文

## 基于非线性回归的 $PM_{2.5}$浓度预报模型

院（系）名 称：数学与统计学院

专 业 名 称 ：统计学

学 生 姓 名 ：曾 婧 鸿

指 导 教 师 ：陈玉蓉 讲师

二〇一八年五月

# 郑 重 声 明

本人呈交的学位论文，是在导师的指导下，独立进行研究工作所取得的成果，所有数据、图片资料真实可靠。尽我所知，除文中已经注明引用的内容外，本学位论文的研究成果不包含他人享有著作权的内容。对本论文所涉及的研究工作做出贡献的其他个人和集体，均已在文中以明确的方式标明。

本人签名：________________　　　　日期：________________

# 摘　要

$PM_{2.5}$浓度预报对治理武汉市大气污染有重大意义。本文基于非线性回归提出了$PM_{2.5}$浓度预报模型，包括单值预报模型与区间预报模型。单值预报模型能较准确预报次日$PM_{2.5}$浓度，在拟合分析中预报偏差约为 6 μg/ $m^3$。区间预报模型可有效预报高浓度与低浓度天气，在模型检验中能覆盖60%-80%观测样本。同时本文结合 NCEP CFS2 气象预报系统实现了$PM_{2.5}$浓度预报模型的预报应用，并建立了 NCEP CFS2 的$PM_{2.5}$浓度预报模型以提高预报准确度，体现了$PM_{2.5}$浓度预报模型的独立预报能力。

# ABSTRACT

Forecasting $PM_{2.5}$ concentration is important to solving air pollution problems in Wuhan. This paper proposes a $PM_{2.5}$ concentration forecast model based on nonlinear regression, including a single-value forecast model and an interval forecast model. The single-value forecast model can precisely forecast $PM_{2.5}$ concentration for the next day, with forecast bias about 6 μg / $m^3$ in goodness of fit analysis. The interval forecast model can efficiently forecast high-concentration and low-concentration days, which covers 60%-80% observed samples in model validation. Moreover, this paper combines the $PM_{2.5}$ concentration forecast model with NCEP Climate Forecast System Version 2 to realize its forecast application, then develops NCEP CFS2's $PM_{2.5}$ concentration forecast model to enhance forecast accuracy. The results indicate that the $PM_{2.5}$ concentration forecast model has good capacity for independent forecasting.

# 目　　录

# 1　关于$PM_{2.5}$浓度预报

## 1.1 实际背景

当下，全国众多城市存在不同程度的空气污染，空气状况的明显恶化不仅危害着人们的健康与正常生活，也阻碍社会经济、卫生等多方事业的发展。影响空气质量的主要因素有人为排放、地形地貌、气象气候等，直接反映于空气污染物浓度的高低。空气污染物主要包括有害气体与颗粒物，而颗粒物正逐渐成为全国大部分城市的首要空气污染物。

目前，随着城市化进程加快，武汉市的大气污染问题愈发严重，$PM_{2.5}$成为武汉市全年大部分时间段内的首要污染物。$PM_{2.5}$是粒径小于等于2.5微米的颗粒物，可被人体直接吸入，因此对人体健康危害极大。$PM_{2.5}$与气象因素之间的相互作用极其复杂，其浓度也常受气象变化影响。

在近几年内，$PM_{2.5}$通常在冬季及前后达到浓度高峰，且可与臭氧形成复合型污染，极大恶化了武汉市大气环境，也严重破坏人体呼吸系统、循环系统等正常生理机能，进而引发哮喘、支气管炎和心血管方面的疾病。为了提高武汉市空气质量、推动社会卫生环保事业发展、保护人体健康安全，开展关于$PM_{2.5}$的研究是解决空气污染问题的有效且必需途径之一。同时随着公众愈发关注因$PM_{2.5}$引发的各类问题，对$PM_{2.5}$浓度的实时信息和预报信息的需求将会日益增加。

## 1.2 研究背景

为最大化获取空气污染物浓度信息，关于空气污染物浓度预报模型[1]的研究已有诸多探索。前人的研究主要采用多元线性回归模型(MLR)、自回归移动平均模型（ARIMA）、人工神经网络模型（ANN）等[4,11]。后有学者不断提出新方法，如混合模型(Hybrid)、非线性回归模型(NLR)，模糊理论模型(Fuzzy)等[2,3,9-10]。这些方法已成功应用于局部地区关于$PM_{2.5}$、臭氧、空气质量的研究中。

考虑到空气污染物浓度易受气象影响，前人的研究基本以气象因素为突破口，探究空气污染物浓度与气象因素的关系，以期建立较为完善的模型并用于公开业务或环境评估。其中，部分关于$PM_{2.5}$的研究结果表明各地域地理因素对$PM_{2.5}$浓度影响极大，因此关于$PM_{2.5}$的研究呈现出个体性，需单独分析不同地域的特性，并

兼顾多类气象因素的综合作用。

关于$PM_{2.5}$浓度预报模型的研究各有特点，或进行多源数据分析或采用案例对比，且理论方法纷杂，没有统一性；部分研究提出了一些新颖且有效的算法，但实现难度大，普适性不足。其中，Cobourn 在 2010 年提出使用非线性回归建立$PM_{2.5}$浓度预报模型[5]。他基于其之前关于臭氧浓度非线性预报模型的研究[6-8]并将相关方法延伸于$PM_{2.5}$，展现了非线性回归对于此类问题的适用性，其思想值得采纳与借鉴；但 Cobourn 建立的预报模型仅适用于特定地区，其参考价值有限。

目前中国尚未有发展成熟的$PM_{2.5}$浓度预报模型，也未有基于武汉市的相关成熟模型，其研究成果也极为有限，充足的研究空间保留着该问题的研究潜力。

## 1.3 研究思路

### 1.3.1 研究内容

鉴于实际背景与研究背景所述：武汉市的首要空气污染物$PM_{2.5}$具有充分的研究价值、非线性回归模型有很好的应用潜力、同时关于该污染物的研究能够带来显著的社会效益并满足公众对于相关信息的实际需求，本文以武汉市为研究地点，以$PM_{2.5}$为研究对象，以各气象因素为研究变量，以非线性回归为研究方法，旨在建立关于$PM_{2.5}$浓度的预报模型，探究$PM_{2.5}$浓度与各气象因素之间的关系，并为武汉市治理$PM_{2.5}$大气污染与监控环境空气质量提供有效依据。

研究内容顺次包括基于非线性回归建立$PM_{2.5}$单值预报模型，建立$PM_{2.5}$区间预报模型，检验$PM_{2.5}$浓度预报模型，与特定气象预报系统结合以实现$PM_{2.5}$浓度预报模型的预报应用。其中，$PM_{2.5}$单值预报模型与$PM_{2.5}$区间预报模型共同组成$PM_{2.5}$浓度预报模型。

建立相对准确可靠的单值预报模型是建立合理有效的区间预报模型之基础。建立$PM_{2.5}$单值预报模型指建立体现$PM_{2.5}$浓度与各气象因素之间关系的模型，能较准确刻画$PM_{2.5}$浓度的实际变化趋势。建立$PM_{2.5}$区间预报模型指在$PM_{2.5}$单值预报模型的基础上建立合理的预报区间，进一步提高预报效果。实现预报应用指对$PM_{2.5}$浓度预报模型引入来自特定气象预报系统的预报数据并分析$PM_{2.5}$浓度预报模型的预报效果、预报特点与独立预报能力。独立预报指$PM_{2.5}$浓度预报模型在仅知预报变量预报数据时可以直接预报次日$PM_{2.5}$浓度，而不需要前一天$PM_{2.5}$浓度信息。

建立$PM_{2.5}$浓度预报模型使用建模数据集，检验模型使用验模数据集，预报应用使用预报数据集。

### 1.3.2 原始数据说明

研究所用数据共有三部分，分别是 2014 年至 2017 年全四年武汉市$PM_{2.5}$浓度逐日观测值，2014 年至 2017 年全四年武汉市气象因素逐日观测值，2016 年 10 月至 12 月与 2017 年全年武汉市气象因素逐日预报值。

第一部分数据是响应变量数据集，来自中国空气质量在线监测分析平台 https://www.aqistudy.cn/historydata/daydata.php?city=%E6%AD%A6%E6%B1%89。该平台直接提供的$PM_{2.5}$浓度逐日数据来自中国环境监测总站。

第二部分数据是预报变量观测值数据集，来自中国气象数据网，网址是 http://data.cma.cn/data/cdcdetail/dataCode/SURF_CLI_CHN_MUL_DAY_CES_V3.0.html。数据经过质量控制，实有率普遍在 99%以上，正确率接近 100%。

第三部分数据是预报变量预报值数据集，全称 NCEP Climate Forecast System Version 2 (CFSv2) Selected Hourly Time-Series Products，来自 Research Data Archive at the National Center for Atmospheric Research, Computational and Information Systems Laboratory，网址是 https://doi.org/10.5065/D6N877VB，对应的气象预报系统是 NCEP Climate Forecast System Version 2 (CFSv2)。

建模数据集指 2014 年至 2016 年全三年$PM_{2.5}$浓度观测值与气象因素观测值数据，用于建立模型。验模数据集指 2017 年全年$PM_{2.5}$浓度观测值与气象因素观测值数据，用于检验模型。预报数据集指 2016 年 10 月至 12 月与 2017 年全年$PM_{2.5}$浓度观测值与气象因素逐日预报值数据，用于实现预报应用。

### 1.3.3 数据预处理

从三类数据集中截选每年 1 月至 3 月与 10 月至 12 月的逐日数据。见附录 A。

建模数据集中存在三个缺失数据的观测样本，予以直接剔除。其中，降水量变量介于 0-0.1mm 之间时被记录为“微量”，无数值型数据。考虑到 0-0.1mm 降水量接近于 0，故直接对此部分数据取 0 处理。各变量数据间数量级相近，且为保持数据整数结构以减小计算过程中产生的舍入误差，不采用中心标准化处理数据。

建模数据集中采用的气象自变量总共 7 个，分别是平均气温、最低气温、最高气温、最小相对湿度、20 至 20 时降水量、蒸发量、最大风速。为有效利用温度变量的信息，此处构造一个温差变量以探究其与$PM_{2.5}$浓度的关系。温差定义为最高气温与最低气温之差。在考虑温差变量时暂不考虑最低气温与最高气温。

$PM_{2.5}$浓度记为$PM_{2.5}$因变量，气象变量记为相应的自变量。各变量单位见表 1.1。后文默认各变量单位不变，故不逐一标注。

表 1.1 **$PM_{2.5}$因变量与部分气象自变量信息表**

| 变量 | $PM_{2.5}$ | 温度 | 温差 | 风速 | 降水量 | 蒸发量 | 相对湿度 |
|---|---|---|---|---|---|---|---|
| 记号 | *pm* | *t* | *trg* | *w* | *pc* | *ep* | *hm* |
| 单位 | 1μg/m$^3$ | 0.1℃ | 0.1℃ | 0.1m/s | 0.1mm | 0.1mm | 1% |
| 全称 | – | 平均气温 | – | 最大风速 | 20 至 20 时降水量 | 大型蒸发量 | 最小相对湿度 |

# 2　基于非线性回归的 $PM_{2.5}$浓度预报模型

## 2.1 非线性回归理论

本文采用 Bates 曲率度量判断非线性回归模型的非线性强度，作为诊断非线性回归模型的一重标准；采用 Box 偏差度量初步判断非线性回归模型参数估计的合理性，作为诊断模型的二重标准；并结合线性近似理论构造非线性回归模型的预报区间。线性近似、Bates 曲率度量、Box 偏差度量等理论来自文献[12]，作者对文献理论进行了整理与总结。程序说明见附录 B。

### 2．1．1 非线性回归模型

非线性回归模型表示为

$$y_i = f(\boldsymbol{x}_i,\boldsymbol{\theta}) + \varepsilon_i, i = 1, 2, ..., n \tag{2.1}$$

其中，随机变量 $y_i$ 是第 $i$ 个响应变量； $f$ 是期望函数； $\boldsymbol{x}_i = (x_{i1}, x_{i2}, ..., x_{ip})'$ 是来自 $n \times p$ 阶样本资料阵 X 的第 $i$ 个样本， $p$ 表示自变量个数， $n$ 表示样本量；

$\boldsymbol{\theta} =(\theta_1, \theta_2, ..., \theta_q)'$ 表示未知参数， $q$ 表示未知参数个数； $\varepsilon_{\mathrm{i}}$ 表示随机误差项。非线性回归模型中至少存在一个参数，满足期望函数关于该参数的导数至少与某个参数有关。

非线性回归模型的基本假设有：(1) $\mathrm{E}(\varepsilon_i) = 0$ ， $\mathrm{var}(\varepsilon_i) = \sigma^2$， $i = 1, 2, ..., n$ ；(2) $\mathrm{cov}(\varepsilon_i, \varepsilon_j) = 0$， $i \neq j$ 。进一步，当 $\varepsilon_i \sim \mathrm{N}(0, \sigma^2)$ 时满足正态假设。

### 2．1．2 线性近似

在固定参数值 $\boldsymbol{\theta}_0$ 处对期望函数进行一阶泰勒展开：

$$f(\boldsymbol{x}_i,\boldsymbol{\theta}) = f(\boldsymbol{x}_i,\boldsymbol{\theta}) + \sum_{j=1}^{q}(\theta_j - \theta_{j0})\ \ \partial f(\boldsymbol{x}_i,\boldsymbol{\theta}) / \partial\theta_j \Big|_{\theta_{j0}} \tag{2.2}$$

其中 $i = 1, 2, ..., n$ 。将此式记为向量形式，则有：

$$f(\mathrm{X},\boldsymbol{\theta}) = f(\mathrm{X},\boldsymbol{\theta}_0) + \partial f(\mathrm{X},\boldsymbol{\theta}) / \partial\boldsymbol{\theta}' \Big|_{\boldsymbol{\theta}_0} (\boldsymbol{\theta} - \boldsymbol{\theta}_0) \tag{2.3}$$

其中， $f(\mathrm{X},\boldsymbol{\theta}) = \left(f(\boldsymbol{x}_1,\boldsymbol{\theta}), f(\boldsymbol{x}_2,\boldsymbol{\theta}), ..., f(\boldsymbol{x}_n,\boldsymbol{\theta})\right)'_{n\times 1}$， $f(\mathrm{X},\boldsymbol{\theta}_0)$ 同理。该形式说明真实期望曲面可以由 $f(\mathrm{X},\boldsymbol{\theta}_0)$ 处的切平面线性近似表示，同时该切平面符合均匀坐

标系并且足够平坦。因此推出线性近似的两个假设条件：（1）平坦性假设:切平面近似期望平面；（2）均匀坐标假设:切平面坐标系近似线性坐标系。

使用线性近似需满足两个假设条件：平坦性假设与均匀坐标假设。判断非线性回归模型是否满足两个假设条件需对模型的非线性强度进行合适度量。

### 2.1.3 Bates 曲率度量

Bates 曲率度量反映非线性回归模型的非线性强度。非线性回归模型可以理解为$q$维参数空间到$n$维响应空间的映射。经过参数空间固定点$\boldsymbol{\theta}_0$处的直线被映射为经过响应空间固定点$f(\mathrm{X},\boldsymbol{\theta}_0)$处的曲线。

假设经过参数空间固定点$\boldsymbol{\theta}_0$处的直线表示为：

$$\boldsymbol{\theta}_b = \boldsymbol{\theta}_0 + b\mathbf{h} \tag{2.4}$$

其中，$b$为常数，$\mathbf{h}$为$q$维非零向量。将期望函数$f(\mathrm{X},\boldsymbol{\theta})$简记为$f(\boldsymbol{\theta})$，得到响应空间对应固定点的曲线：

$$f\left(\boldsymbol{\theta}_b\right) = f\left(\boldsymbol{\theta}_0 + b\mathbf{h}\right) \tag{2.5}$$

记$\partial f\left(\boldsymbol{\theta}_0\right)/\partial\boldsymbol{\theta}_0{}' = \dot{\mathrm{V}}$。在$b=0$处$f\left(\boldsymbol{\theta}_b\right)$的切线斜率为：

$$f\,'\left(\boldsymbol{\theta}_b\right) = \left.\frac{df\left(\boldsymbol{\theta}_b\right)}{db}\right|_{b=0} = \left.\frac{\partial f\left(\boldsymbol{\theta}_b\right)}{\partial\boldsymbol{\theta}_b{}'}\frac{d\boldsymbol{\theta}_b}{db}\right|_{b=0} \doteq \underset{n\times q}{\dot{\mathbf{V}}}\;\underset{q\times 1}{\mathbf{h}} \tag{2.6}$$

记$\partial\dot{\mathrm{V}}/\partial\boldsymbol{\theta}_0{}' = \ddot{\mathrm{V}}$。在$b=0$处$f\left(\boldsymbol{\theta}_b\right)$的二阶导为：

$$f\,''\left(\boldsymbol{\theta}_b\right) = \left.\frac{d^2 f\left(\boldsymbol{\theta}_b\right)}{db^2}\right|_{b=0} = \left.\frac{d\left(\dot{\mathbf{V}}\mathbf{h}\right)}{db}\right|_{b=0} \doteq \underset{1\times q}{\mathbf{h}'}\;\underset{n\times q\times q}{\ddot{\mathbf{V}}}\;\underset{q\times 1}{\mathbf{h}} \tag{2.7}$$

将该二阶导看作切线斜率的“加速度”，可以分解为相互垂直的两向量:

$$f\,''\left(\boldsymbol{\theta}_b\right) = f_1{}''\left(\boldsymbol{\theta}_b\right) + f_2{}''\left(\boldsymbol{\theta}_b\right) \tag{2.8}$$

Bates 与 Watts 分别定义固有曲率（intrinsic curvature）与参数效应曲率（parameter-effects curvature）。固有曲率表示非线性模型本身的非线性强度，由模型中参数与自变量的组合形式决定，记为$\mathrm{K}_{\mathbf{h}}^{\mathrm{N}}$。参数效应曲率表示非线性参数引起的非线性强度，可以通过重新参数化降低,记为$\mathrm{K}_{\mathbf{h}}^{\mathrm{P}}$。定义：

$$\mathrm{K}_{\mathbf{h}}^{\mathrm{N}} = \left\|f_1{}''\left(\boldsymbol{\theta}_b\right)\right\| / \left\|f\,'\left(\boldsymbol{\theta}_b\right)\right\|^2 \tag{2.9}$$

$$\mathrm{K}_{\mathbf{h}}^{\mathrm{P}} = \left\|f_2{}''\left(\boldsymbol{\theta}_b\right)\right\| / \left\|f\,'\left(\boldsymbol{\theta}_b\right)\right\|^2 \tag{2.10}$$

当固有曲率很小时模型满足平坦性假设，当参数效应曲率很小时模型满足均匀坐标假设。

为方便计算固有曲率与参数效应曲率，采用正交变换旋转样本空间以分解二阶导立体阵。首先将一阶导矩阵进行$QR$分解：

$$\underset{n\times q}{\dot{\mathbf{V}}}=\underset{n\times n}{\mathbf{Q}}\,\underset{n\times q}{\mathbf{R}}=\left(\mathbf{Q}_1\middle|\mathbf{Q}_2\right)\left(\frac{\mathbf{R}_1}{\mathbf{0}}\right)=\underset{n\times q}{\mathbf{Q}_1}\,\underset{q\times q}{\mathbf{R}_1}\tag{2.11}$$

记$\underset{q\times q}{\mathbf{L}}=\mathbf{R}_1^{-1}$。再进行坐标变换：

$$\Phi=\mathbf{R}_1\left(\boldsymbol{\theta}-\hat{\boldsymbol{\theta}}\right)\tag{2.12}$$

$$\boldsymbol{\theta}=\hat{\boldsymbol{\theta}}+\mathbf{L}\Phi\tag{2.13}$$

当$\Phi=0$时$\boldsymbol{\theta}=\hat{\boldsymbol{\theta}}$。在$\Phi=0$处对期望函数$f\left(\boldsymbol{\theta}\right)$关于$\Phi$求导：

$$\dot{\mathrm{M}}=\left.\frac{\partial f\left(\boldsymbol{\theta}\right)}{\partial\Phi}\right|_{\Phi=0}=\left.\frac{\partial f\left(\boldsymbol{\theta}\right)}{\partial\boldsymbol{\theta}'}\frac{\partial\boldsymbol{\theta}}{\partial\Phi}\right|_{\Phi=0}=\dot{\mathbf{V}}\mathbf{L}=\mathbf{Q}_1\tag{2.14}$$

$$\ddot{\mathrm{M}}=\left.\frac{\partial\dot{\mathrm{M}}}{\partial\Phi}\right|_{\Phi=0}=\mathbf{L}'\ddot{\mathbf{V}}\mathbf{L}\tag{2.15}$$

进行坐标变换后的一阶导矩阵$\dot{\mathbf{V}}$变为$\dot{\mathrm{M}}$，二阶导立体阵$\ddot{\mathbf{V}}$变为$\ddot{\mathrm{M}}$，故可直接对$\ddot{\mathrm{M}}$分析非线性强度。

进行正交变换，对二阶导样本空间进行旋转：

$$\underset{n\times q\times q}{\ddot{\mathrm{A}}}\doteq\underset{n\times n}{\mathbf{Q}}'\underset{n\times q\times q}{\ddot{\mathrm{M}}}=\left(\underset{n\times q}{\mathbf{Q}_1}\middle|\underset{n\times(n-q)}{\mathbf{Q}_2}\right)'\ddot{\mathrm{M}}=\left(\frac{\mathbf{Q}_1'\ddot{\mathrm{M}}}{\mathbf{Q}_2'\ddot{\mathrm{M}}}\right)\doteq\left(\frac{\ddot{\mathrm{A}}^{\mathrm{P}}}{\ddot{\mathrm{A}}^{\mathrm{N}}}\right)\tag{2.16}$$

$\ddot{\mathrm{M}}$与$\ddot{\mathrm{A}}$表示三维立体阵，各包含$n$个面，每个面是一个$q\times q$阶矩阵。$\underset{q\times q\times q}{\ddot{\mathrm{A}}^{\mathrm{P}}}$含有$q$面，$\underset{(n-q)\times q\times q}{\ddot{\mathrm{A}}^{\mathrm{N}}}$含有$n-q$面，每个面均代表一个样本。

$\mathbf{Q}_1$与$\mathbf{Q}_2$空间中的向量相互正交。仅考虑样本空间维数时得到$\mathbf{Q}_1$作用的$q$维空间$\underset{q\times q\times q}{\ddot{\mathrm{A}}^{\mathrm{P}}}$与$\mathbf{Q}_2$作用的$n-q$维$\underset{(n-q)\times q\times q}{\ddot{\mathrm{A}}^{\mathrm{N}}}$空间，这两个空间相互正交，由此原样本空间$\ddot{\mathbf{V}}$被分解为相互正交的两个空间$\ddot{\mathrm{A}}^{\mathrm{P}}$与$\ddot{\mathrm{A}}^{\mathrm{N}}$。

从$\dot{\mathbf{V}}$的定义可知，$\dot{\mathbf{V}}$由$q$维参数空间生成，而从$\dot{\mathbf{V}}$的$QR$分解可知，$\dot{\mathbf{V}}$为$\mathbf{Q}$的前$q$维正交基生成的空间，因此$q$维参数空间对应$\mathbf{Q}_1$生成的空间，参数空间引起

的参数效应通过$\mathbf{Q}_1$传递至正交变换后的空间。故有$q$维空间$\ddot{\mathrm{A}}^{\mathrm{P}}$对应参数空间，表示产生参数效应非线性的空间，而该空间的正交补空间$\ddot{\mathrm{A}}^{\mathrm{N}}$则表示产生固有非线性的空间。

取$\mathbf{h}=\mathbf{L}d$，$d=\left(d_1,d_2,...,d_q\right)'$满足$\|d\|=1$，则根据固有曲率、参数效应曲率公式得出：

$$f'\left(\boldsymbol{\theta}_b\right)=\dot{\mathbf{V}}\mathbf{h}=\dot{\mathbf{V}}\mathbf{L}d=\mathbf{Q}_1 d \qquad (2.17)$$

$$f''\left(\boldsymbol{\theta}_b\right)=\mathbf{h}'\ddot{\mathbf{V}}\mathbf{h}=d'\mathbf{L}'\ddot{\mathbf{V}}\mathbf{L}d=d'\ddot{\mathrm{M}}d \qquad (2.18)$$

推出：

$$\mathrm{K}_{\mathbf{h}}^{\mathrm{N}}=\left\|\left(d'\ddot{\mathrm{M}}d\right)^{\mathrm{N}}\right\|=\left\|d'\ddot{\mathrm{A}}^{\mathrm{N}}d\right\| \qquad (2.19)$$

$$\mathrm{K}_{\mathbf{h}}^{\mathrm{P}}=\left\|\left(d'\ddot{\mathrm{M}}d\right)^{\mathrm{P}}\right\|=\left\|d'\ddot{\mathrm{A}}^{\mathrm{P}}d\right\| \qquad (2.20)$$

$\mathrm{K}_{\mathbf{h}}^{\mathrm{N}}$和$\mathrm{K}_{\mathbf{h}}^{\mathrm{P}}$与单位向量$d$的取值有关，故考虑所有方向上的均方曲率作为统一评价标准。定义均方固有曲率（Mean Square Intrinsic Curvature）与均方参数效应曲率（Mean Square Parameter-effects Curvature）分别为$\mathrm{K}_{MS}^{\mathrm{N}}$与$\mathrm{K}_{MS}^{\mathrm{P}}$：

$$\left(\mathrm{K}_{MS}^{\mathrm{N}}\right)^2=\left(\int_{\|d\|=1}\sum_{i=1}^{n-q}\left(d'\ddot{\mathrm{A}}_i^{\mathrm{N}}d\right)^2 ds\right)\Big/S\left(q\right) \qquad (2.21)$$

$$\left(\mathrm{K}_{MS}^{\mathrm{P}}\right)^2=\left(\int_{\|d\|=1}\sum_{i=1}^{q}\left(d'\ddot{\mathrm{A}}_i^{\mathrm{P}}d\right)^2 ds\right)\Big/S\left(q\right) \qquad (2.22)$$

其中，$S\left(q\right)$表示$q$维单位球体表面积；$\ddot{\mathrm{A}}_i^{\mathrm{N}}$表示$\ddot{\mathrm{A}}^{\mathrm{N}}$的第$i$个样本面，$\ddot{\mathrm{A}}_i^{\mathrm{P}}$同理。

将曲面积分展开：

$$\left(\mathrm{K}_{MS}^{\mathrm{N}}\right)^2=\sum_{t=1}^{n-q}\left\{2\sum_{i=1}^{q}\sum_{j=1}^{q}{a_{tij}}^2+\left(\sum_{i=1}^{q}a_{tii}\right)^2\right\}\Big/q\left(q+2\right) \qquad (2.23)$$

$$\left(\mathrm{K}_{MS}^{\mathrm{P}}\right)^2=\sum_{t=1}^{q}\left\{2\sum_{i=1}^{q}\sum_{j=1}^{q}{a_{tij}}^2+\left(\sum_{i=1}^{q}a_{tii}\right)^2\right\}\Big/q\left(q+2\right) \qquad (2.24)$$

其中，$a_{tij}$表示对应立体阵的第$t$个样本面的$\left(i,j\right)$元。

$\mathrm{K}_{MS}^{\mathrm{N}}$用于衡量模型固有非线性强度，对应平坦性假设。$\mathrm{K}_{MS}^{\mathrm{P}}$用于衡量模型参数效应非线性强度，对应均匀坐标假设。由于线性回归模型的$\ddot{\mathbf{V}}$恒为0，要对非线性回归模型采用线性近似或证明非线性回归模型的性态充分接近线性性态则必须使得$\mathrm{K}_{MS}^{\mathrm{N}}$与$\mathrm{K}_{MS}^{\mathrm{P}}$足够接近0。因此需要寻找一个临界值以判断均方曲率与0足够接

近，当均方曲率小于此临界值时认为均方曲率接近于 0，此时满足相应的假设条件。临界值选取方式不唯一，下面提供一种选取方式。

考虑$n\times q$阶样本资料阵 X 的线性回归模型。置信水平为$(1-\alpha)$的参数联合置信域为$(\hat{\boldsymbol{\theta}}-\boldsymbol{\theta})'(\mathrm{X}'\mathrm{X})(\hat{\boldsymbol{\theta}}-\boldsymbol{\theta})\le q\hat{\sigma}^2F(q,n-q,\alpha)$，其中$\hat{\sigma}$是残差标准差。该置信域对应圆面半径为$\hat{\sigma}\sqrt{qF(q,n-q,\alpha)}$。由于曲率的倒数为曲率半径，此时线性置信域的曲率为$1\big/\hat{\sigma}\sqrt{qF(q,n-q,\alpha)}$。可取该线性置信域曲率为临界值。

记$\rho=\hat{\sigma}\sqrt{q}$。使用线性近似时需满足：

$$\mathrm{K}_{MS}^{\mathrm{N}}\le 1\big/\rho\sqrt{F(q,n-q,\alpha)} \tag{2.25}$$

$$\mathrm{K}_{MS}^{\mathrm{P}}\le 1\big/\rho\sqrt{F(q,n-q,\alpha)} \tag{2.26}$$

变形可得$\rho\mathrm{K}_{MS}^{\mathrm{N}}\le 1\big/\sqrt{F(q,n-q,\alpha)}$，$\rho\mathrm{K}_{MS}^{\mathrm{P}}\le 1\big/\sqrt{F(q,n-q,\alpha)}$。

最后只需分别比较$\rho\mathrm{K}_{MS}^{\mathrm{N}}$、$\rho\mathrm{K}_{MS}^{\mathrm{P}}$与$1\big/\sqrt{F(q,n-q,\alpha)}$以判断模型是否满足平坦性假设与均匀坐标假设。当$\rho\mathrm{K}_{MS}^{\mathrm{N}}$、$\rho\mathrm{K}_{MS}^{\mathrm{P}}$均小于$1\big/\sqrt{F(q,n-q,\alpha)}$时认为非线性回归模型充分接近线性性质。此时$\rho\mathrm{K}_{MS}^{\mathrm{N}}$、$\rho\mathrm{K}_{MS}^{\mathrm{P}}$是无量纲数，不受变量单位影响。

非线性回归模型在满足两个假设条件后可能没有有效的参数估计，故在 Bates 曲率度量模型非线性强度的基础上，还需进一步研究参数估计的优良性。

### 2．1．4 Box 偏差度量

Bates 曲率度量从非线性强度出发度量非线性回归模型的性态，但不表示在曲率接近 0 时参数估计有很好的性质。故进而采用 Box 偏差度量模型的近似参数估计偏差，从而初步判断参数估计的优良性。给出 Box 偏差定义：

$$Bias(\hat{\boldsymbol{\theta}})=\mathrm{E}(\hat{\boldsymbol{\theta}}-\boldsymbol{\theta})=-\frac{\sigma^2}{2}(\dot{\mathbf{V}}'\dot{\mathbf{V}})^{-1}\sum_{i=1}^{n}\dot{\mathbf{V}}_i'tr\left((\dot{\mathbf{V}}'\dot{\mathbf{V}})^{-1}\ddot{\mathbf{V}}_i\right) \tag{2.27}$$

其中，$\dot{\mathbf{V}}_i$表示$\dot{\mathbf{V}}$第$i$个$1\times q$阶行向量或样本，$\ddot{\mathbf{V}}_i$表示$\ddot{\mathbf{V}}$第$i$个样本面。

$$\dot{\mathbf{V}}'\dot{\mathbf{V}}=\mathbf{R}'\mathbf{Q}'\mathbf{Q}\mathbf{R}=\mathbf{R}_1'\mathbf{R}_1=(\mathbf{L}\mathbf{L}')^{-1} \tag{2.28}$$

$$tr\left((\dot{\mathbf{V}}'\dot{\mathbf{V}})^{-1}\ddot{\mathbf{V}}_i\right)=tr(\mathbf{L}\mathbf{L}'\ddot{\mathbf{V}}_i)=tr(\mathbf{L}'\ddot{\mathbf{V}}_i\mathbf{L})=tr(\ddot{\mathrm{M}}_i) \tag{2.29}$$

由公式(2.28)与(2.29)推出：

$$\mathrm{E}(\hat{\boldsymbol{\theta}}-\boldsymbol{\theta})=-\frac{\sigma^2}{2}\mathbf{L}\mathbf{L}'\sum_{i=1}^{n}\dot{\mathbf{V}}_i'tr(\ddot{\mathrm{M}}_i) \tag{2.30}$$

同时定义 Box 百分偏差为 Box 偏差与对应参数估计的百分比。Box 百分偏差可进一步衡量参数估计的偏差大小。

当 Box 偏差很大时，参数估计通常不优良。但在参数有偏时非线性回归模型可能会有很好的拟合效果。因此 Box 偏差只作为一个初步参考，并不最终决定参数估计的好坏。

## 2.2 单值预报模型

### 2．2．1 非线性构造

非线性回归模型目前尚未有统一适用的变量选择方法。作者主要基于研究对象的实际性质与变量间的相关性检验初步筛选变量，最终由模型效果对变量选择的合理性进行解释。

检验 $PM_{2.5}$ 因变量与各自变量间的相关性。见表 2.1。在显著水平 $\alpha=.05$ 时，所有 $P-$值均小于 $\alpha$，认为 $PM_{2.5}$ 因变量与各自变量之间的相关性显著。其中，温差与降水量对 $PM_{2.5}$ 因变量的影响最为显著，其次是降水量、温度与风速。

**表 2.1 $PM_{2.5}$ 因变量与各自变量的 Spearman 相关性检验表**

| 变量 | $t$ | $trg$ | $w$ | $pc$ | $ep$ | $hm$ |
|---|---|---|---|---|---|---|
| 相关系数 | -.22 | .27 | -.22 | -.26 | -.18 | -.15 |
| $P-$值 | .00 | .00 | .00 | .00 | .00 | .00 |

在 $PM_{2.5}$ 因变量关于温差、降水量、温度、风速图 2.1 中，$PM_{2.5}$ 因变量与各自变量之间没有明显的线性关系，但均有一定的非线性关系，可直观上无法确定函数形式。其中，$PM_{2.5}$ 因变量与温差之间的走势关系相对更加突出，总体上两者呈正相关。故首先探究 $PM_{2.5}$ 因变量与温差之间可能存在的非线性形式。

对 $PM_{2.5}$ 因变量取对数处理。为了减小数量级差异与舍入误差，对 $PM_{2.5}$ 因变量取对数后乘以 10，提高一级数量级。记取对数后的 $PM_{2.5}$ 因变量为对数 $PM_{2.5}$，记号为 $lpm$。

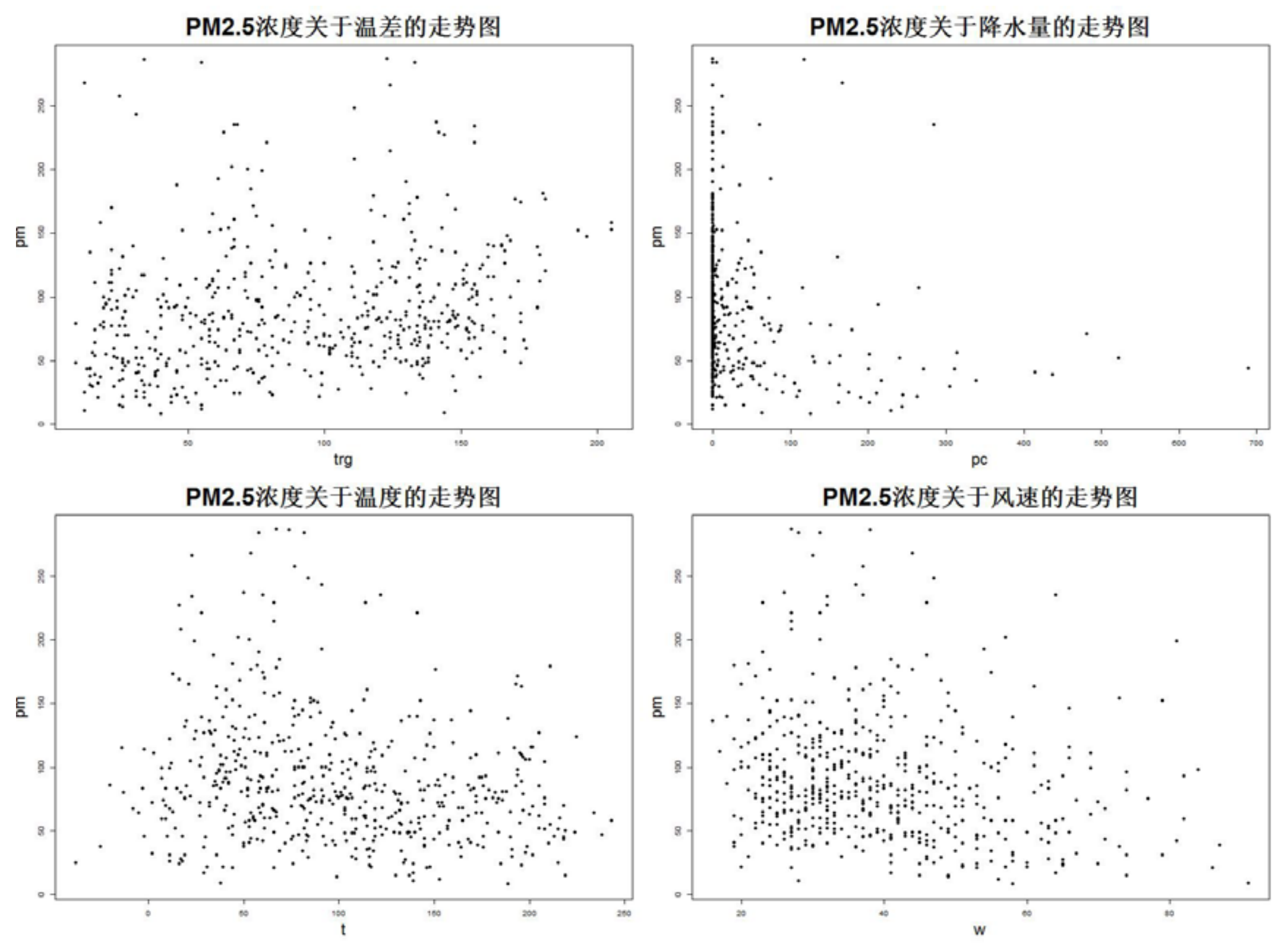

图 2.1 $PM_{2.5}$因变量关于部分自变量的走势图

在对数$PM_{2.5}$与温差图 2.2 中，对数$PM_{2.5}$与温差之间的非线性关系更加明显。使用函数表达式$y = a \times \exp(-b/x)$对两者之间的走势进行刻画，其中$a$、$b$为未知参数。该函数表达式可以较好表示对数$PM_{2.5}$与温差之间的关系，由此构造出非线性回归模型的非线性结构为$lpm = a \times \exp(-b/trg)$ 。

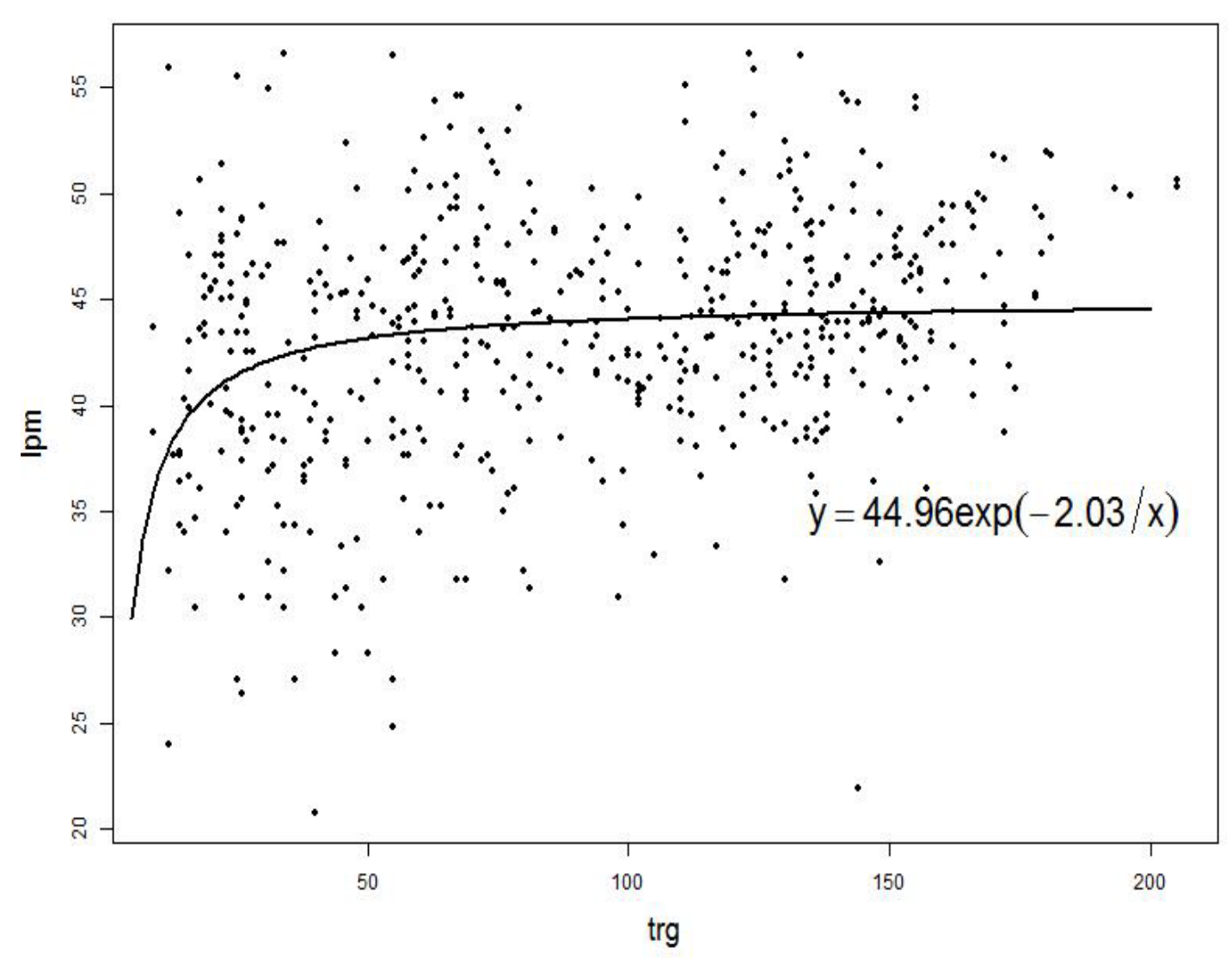

图 2.2 对数$PM_{2.5}$关于温差的非线性走势图

### 2.2.2 初始模型

基于非线性构造结果，对温差变量采用非线性结构，该非线性结构作为模型的主要部分；而将其余变量直接纳入模型以增加模型的解释度。暂不采用 $hm$ 变量。

由此构造出初始模型函数：

$$lpm = \theta_1 \exp\left(-\theta_2 / trg\right) + \theta_3 \times w + \theta_4 \times t + \theta_5 \times pc + \theta_6 \times ep \tag{2.31}$$

初始模型函数(2.31)即为期望函数 $f(\mathrm{X}, \boldsymbol{\theta})$ 的对数形式。

根据模型(2.31)建立非线性回归模型。采用 Gauss-Newton 法求解最小二乘估计量。根据预迭代结果选取初值。迭代信息见表 2.2。其中，迭代 2 步后残差平方和与各参数估计达到稳定，迭代 4 步后确认迭代收敛。

判断初始模型的非线性强度。计算固有曲率与参数效应曲率，并与临界值、临界值的 0.5 倍、临界值的 0.2 倍分别进行比较。见表 2.3。固有曲率与参数效应曲率均小于三种临界值，认为非线性曲率近似为 0，确认初始模型的非线性强度很低并且满足平坦性假设与均匀坐标假设。

同时计算 Box 偏差与 Box 百分偏差，见表 2.4。所有参数估计的偏差很小，百分偏差均小于 0.5%，认为所有参数估计接近无偏估计，初始模型没有产生较大偏差的参数估计，初步确认初始模型的参数估计是合理的。

表 2.2 初始模型迭代信息表

| 步数 | $\hat{\theta}_1$ | $\hat{\theta}_2$ | $\hat{\theta}_3$ | $\hat{\theta}_4$ | $\hat{\theta}_5$ | $\hat{\theta}_6$ | 残差平方和 |
|---|---|---|---|---|---|---|---|
| 4 | 51.201 | 1.828 | -0.062 | -0.011 | -0.013 | -0.147 | 14099.98 |

表 2.3 初始模型非线性强度度量

| 曲率 | 曲率类型 | 曲率值 | 临界值 | 0.5×临界值 | 0.2×临界值 |
|---|---|---|---|---|---|
| $\rho \mathrm{K}_{MS}^{\mathrm{N}}$ | 固有曲率 | 0.007 | 0.688 | 0.344 | 0.138 |
| $\rho \mathrm{K}_{MS}^{\mathrm{P}}$ | 参数效应曲率 | 0.022 | 0.688 | 0.344 | 0.138 |

表 2.4 初始模型参数估计的 **Box** 偏差度量

| 参数估计 | $\hat{\theta}_1$ | $\hat{\theta}_2$ | $\hat{\theta}_3$ | $\hat{\theta}_4$ | $\hat{\theta}_5$ | $\hat{\theta}_6$ |
|---|---|---|---|---|---|---|
| Box 偏差 | 0.002 | 0.003 | 0.00003 | 0.000 | 0.000 | -0.00005 |
| Box 百分偏差 | 0.003% | 0.151% | -0.040% | -0.026% | 0.009% | 0.032% |

进行残差分析。在图 2.3 中，大部分标准化残差均在(-3,3)范围内，少数为疑似异常点或强影响点。在 Spearman 相关性检验中，当显著水平 $\alpha$=.01 时，除了变量 $t$ 与 $ep$，标准化残差绝对值与拟合值、其余变量之间存在显著的相关性，认为标准化残差存在异方差性，故初始模型存在异方差性。

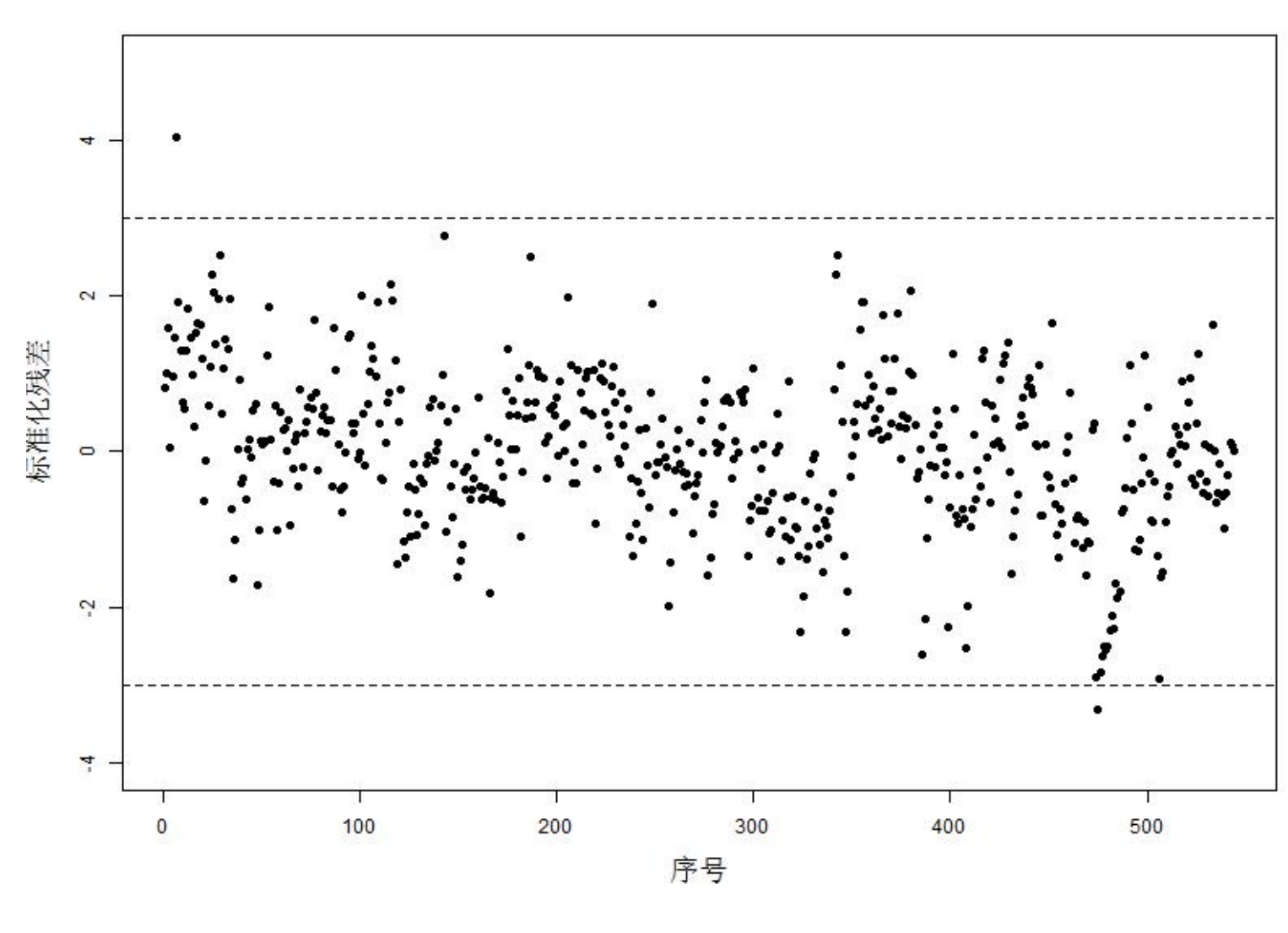

图 2.3 标准化残差图

表 2.5 标准化残差绝对值与拟合值、各自变量的 **Spearman** 相关性检验表

| 变量 | $\hat{y}$ | $t$ | $trg$ | $w$ | $pc$ | $ep$ |
|---|---|---|---|---|---|---|
| 相关系数 | -.142 | -.007 | -.156 | .154 | .168 | -.017 |
| $P-$值 | .0009 | .875 | .0003 | .0003 | .0000 | .687 |

在时序前一个标准化残差 $e_t$ 与时序后一个标准化残差 $e_{t-1}$ 图 2.4 中，标准化残差存在序列正相关性。对时序前后相邻的两组标准化残差做 Pearson 相关性检验，相关系数为.639，$P-$值远小于.01，认为标准化残差存在显著的正自相关性。

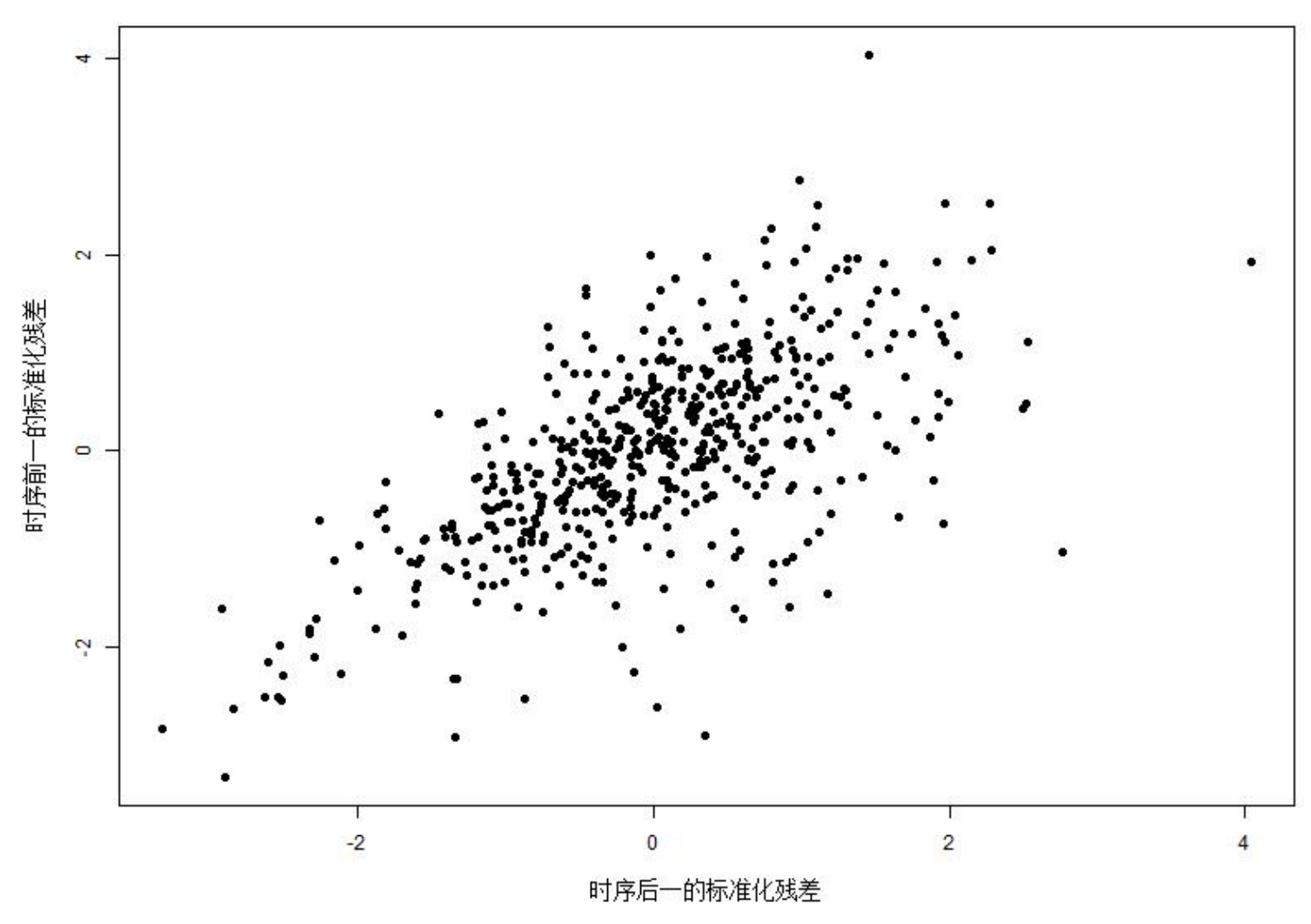

图 2.4 时序前后相邻的标准化残差图

综上所述，初始模型存在异方差性与自相关性，不满足非线性回归模型的基本假设。

### 2．2．3 *id* 模型

为了消弱初始模型的异方差性，首先对建模数据集的对数 $PM_{2.5}$ 进行分析。在图 2.5 中，对数 $PM_{2.5}$ 样本点主要分布于均值附近，部分样本点波动较大。从图 2.6 中，可知对数 $PM_{2.5}$ 分布的两端尾部较厚，且有左侧长尾。大部分对数 $PM_{2.5}$ 样本点集中于均值上下一个标准差范围内的区域，定义该区域为分层界，分层界内样本点方差较小，而分层界上下两区域内的样本点方差较大。

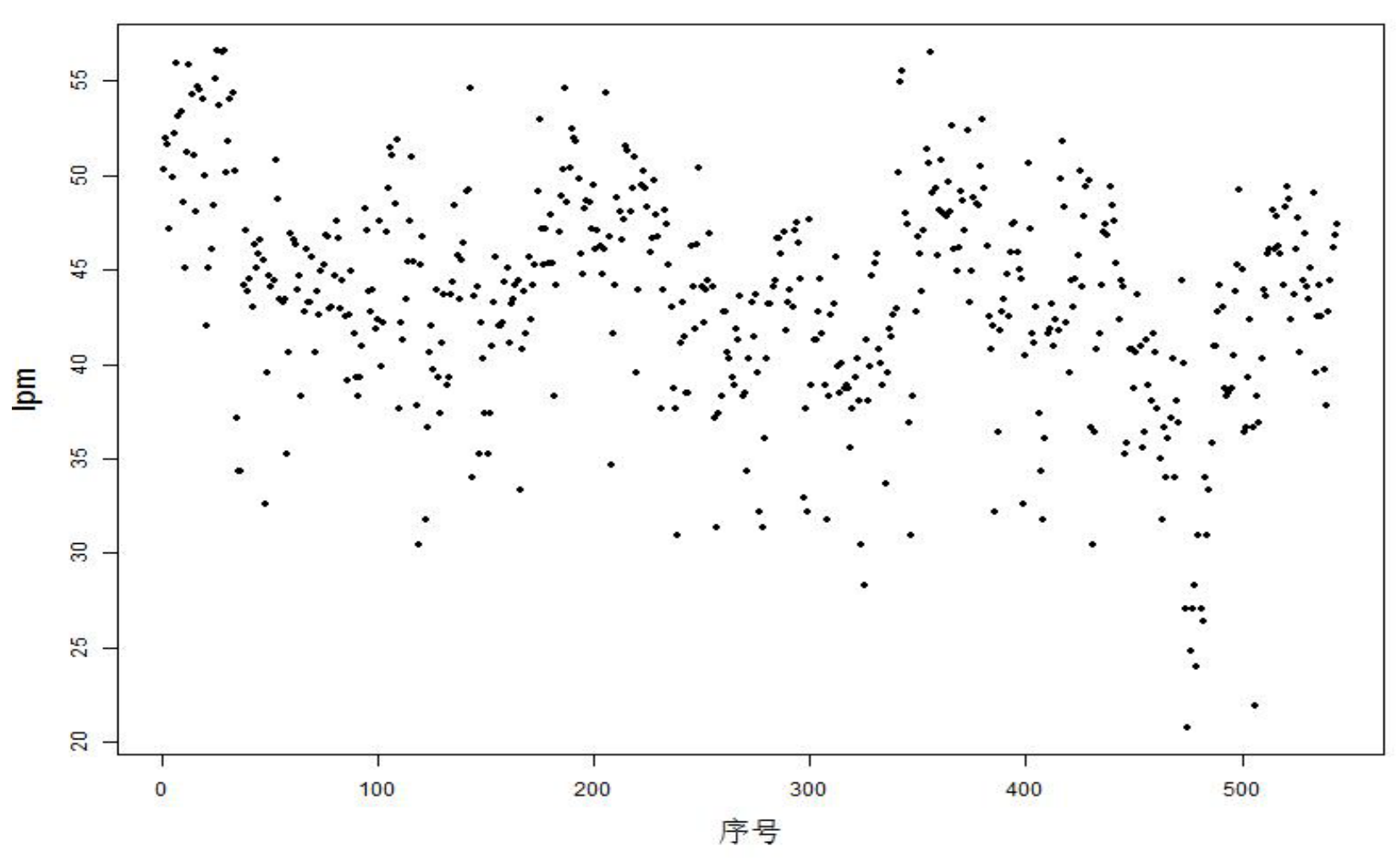

图 2.5 2014 年至 2016 年的对数 $PM_{2.5}$

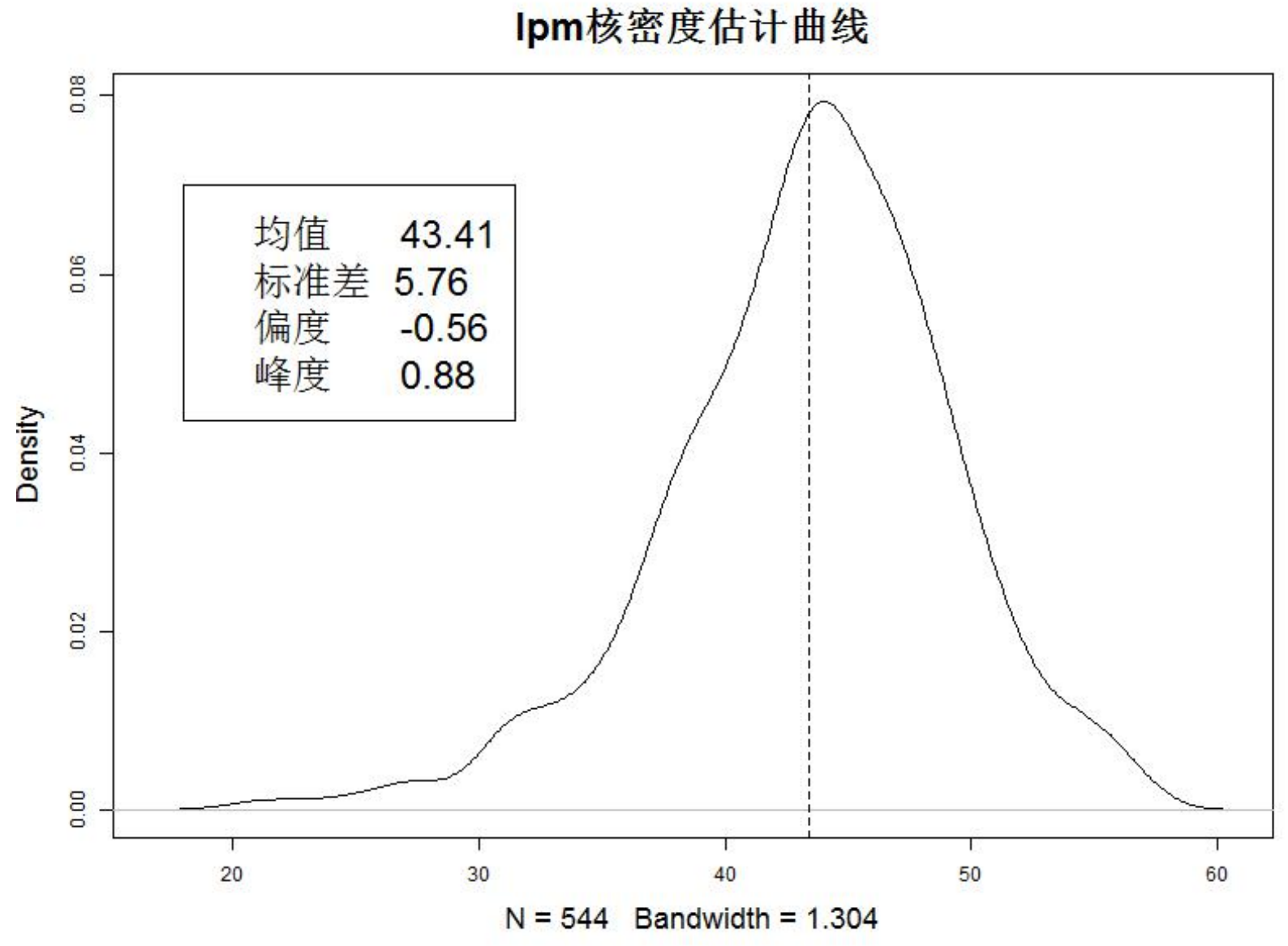

图 2.6 对数 $PM_{2.5}$ 的核密度估计图

依据对数 $PM_{2.5}$ 的性质，对对数 $PM_{2.5}$ 均值加减一个标准差的区间端点值取整。由此构造属性变量 $id$：

$$id=\begin{cases}-1, & lpm\le 35\\ 0, & 35<lpm\le 50\\ 1, & lpm>50\end{cases} \tag{2.32}$$

由此构造 $id$ 模型函数为：

$$lpm=\theta_1\times\exp\left(-\theta_2/trg\right)+\theta_3\times w+\theta_4\times t+\theta_5\times pc+\theta_6\times ep+\theta_7\times id \tag{2.33}$$

根据模型(2.33)建立非线性回归模型。迭代 3 次后确认迭代收敛。固有曲率与参数效应曲率均小于三种临界值，认为曲率近似为 0，确认该模型非线性强度很低。同时，所有参数估计的 Box 偏差很小，百分偏差均小于 0.5%，认为参数估计较为合理。

表 2.6 $id$ 模型迭代信息表

| 步数 | $\hat{\theta}_1$ | $\hat{\theta}_2$ | $\hat{\theta}_3$ | $\hat{\theta}_4$ | $\hat{\theta}_5$ | $\hat{\theta}_6$ | $\hat{\theta}_7$ | 残差平方和 |
| --- | --- | --- | --- | --- | --- | --- | --- | --- |
| 3 | 46.706 | 0.696 | -0.032 | -0.008 | -0.008 | -0.045 | 9.645 | 6074.334 |

表 2.7 *id* 模型非线性强度度量

| 曲率 | 曲率类型 | 曲率值 | 临界值 | 0.5×临界值 | 0.2×临界值 |
| --- | --- | --- | --- | --- | --- |
| $\rho \mathrm{K}_{MS}^{\mathrm{N}}$ | 固有曲率 | 0.004 | 0.702 | 0.351 | 0.140 |
| $\rho \mathrm{K}_{MS}^{\mathrm{P}}$ | 参数效应曲率 | 0.015 | 0.702 | 0.351 | 0.140 |

表 2.8 *id* 模型参数估计的 **Box** 偏差度量

| 参数估计 | $\hat{\theta}_1$ | $\hat{\theta}_2$ | $\hat{\theta}_3$ | $\hat{\theta}_4$ | $\hat{\theta}_5$ | $\hat{\theta}_6$ | $\hat{\theta}_7$ |
| --- | --- | --- | --- | --- | --- | --- | --- |
| Box 偏差 | 0.001 | 0.001 | 0.000 | 0.000 | 0.000 | -0.00002 | -0.0002 |
| Box 百分偏差 | 0.002% | 0.157% | -0.034% | -0.014% | 0.006% | 0.054% | -0.002% |

进行残差分析。在图 2.7 中，所有标准化残差均在(-3,3)范围内，没有异常点。在 Spearman 相关性检验中，当显著水平 $\alpha$=.01 时，标准化残差绝对值与拟合值、所有变量之间无显著相关性，初始模型的异方差性被有效减弱。因此可以认为标准化残差的异方差性已被消除，*id* 模型不存在异方差性。

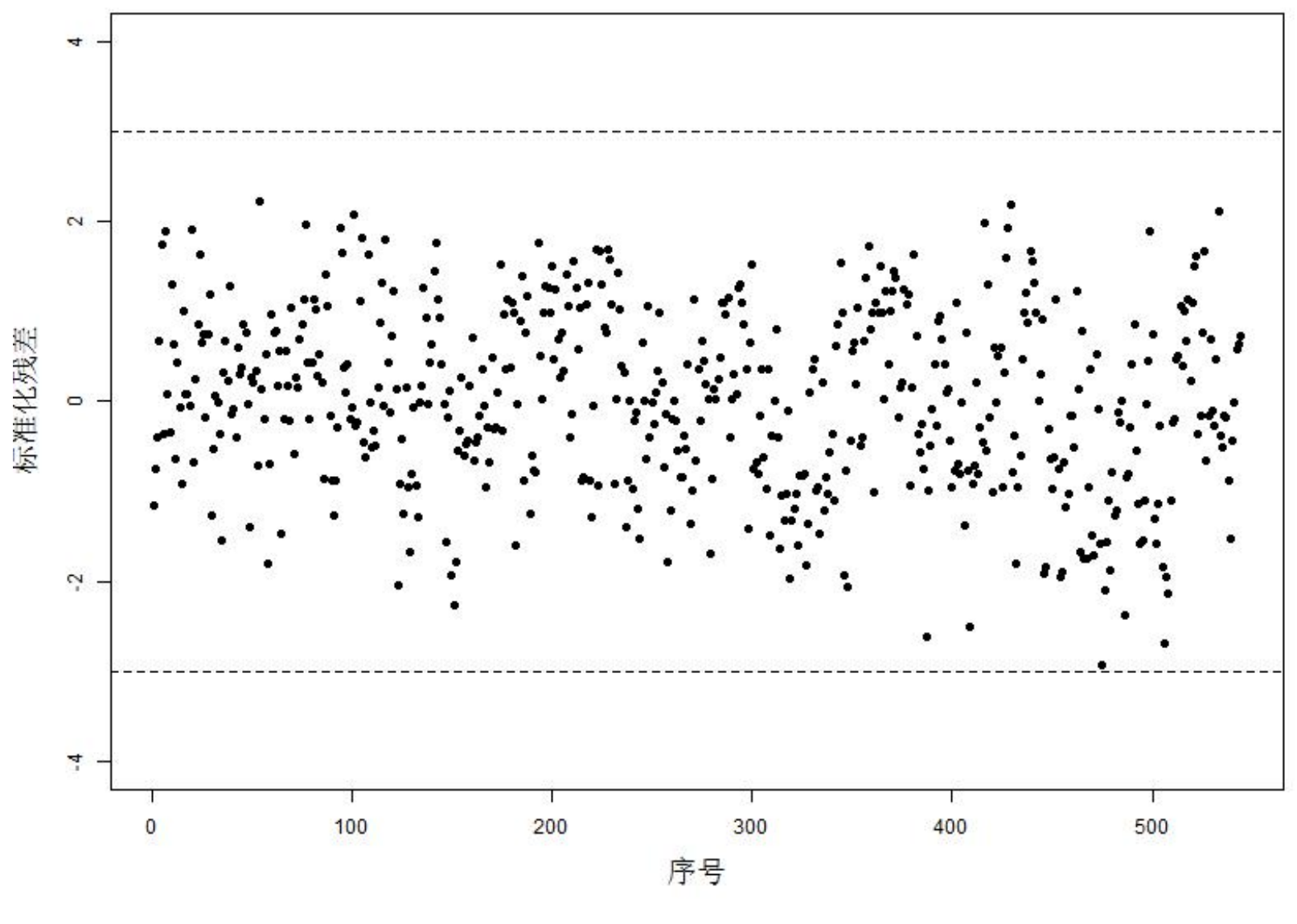

图 2.7 标准化残差图

表 2.9 标准化残差绝对值与拟合值、各自变量的 **Spearman** 相关系数表

| 相关系数 | $\hat{y}$ | $t$ | $trg$ | $w$ | $pc$ | $ep$ | $id$ |
|---|---|---|---|---|---|---|---|
| 初始模型 | -.142 | -.007 | -.156 | .154 | .168 | -.017 | - |
| $id$ 模型 | -.000 | .046 | -.067 | .019 | .098 | -.029 | -.049 |
| $P-$值 | .999 | .289 | .116 | .660 | .022 | .498 | .251 |

在时序前一个标准化残差 $e_t$ 与时序后一个标准化残差 $e_{t-1}$ 图 2.8 中，标准化残差存在序列正相关性。对时序前后相邻的两组标准化残差做 Pearson 相关性检验，相关系数为.417， $P-$值远小于.01，认为标准化残差存在显著的正自相关性。

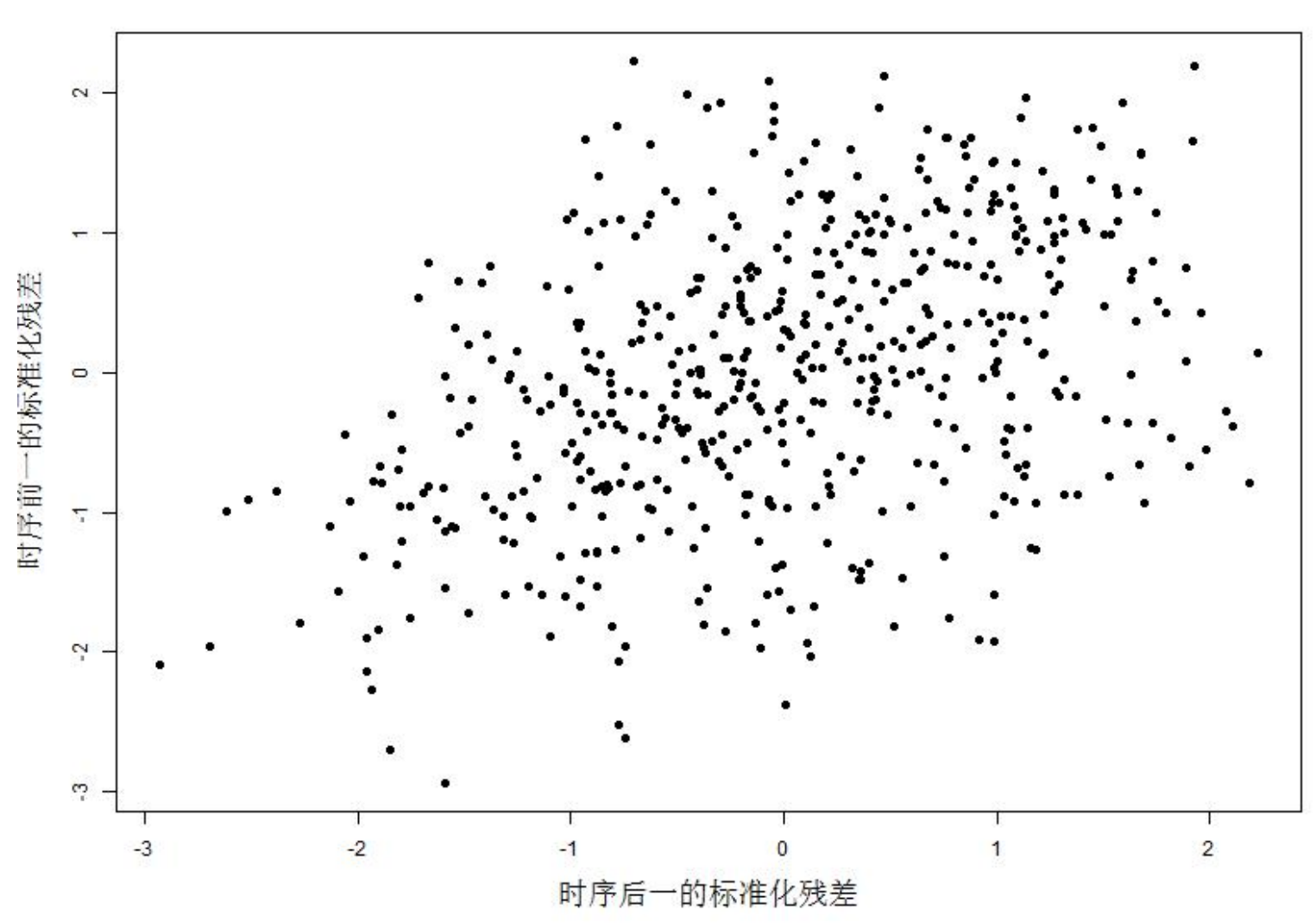

图 2.8 时序前后相邻的标准化残差图

综上所述，$id$ 模型不存在异方差性，而存在自相关性，满足非线性回归模型关于随机误差项零均值、等方差的基本假设，但不满足随机误差不相关的假设。

### 2．2．4 迭代模型

根据上述分析结果，残差之间存在一定的正相关性，记相关系数为 $\rho$ 。采用一阶迭代法消除模型自相关性。假设残差之间存在一阶自回归形式：

$$e_t = \rho e_{t-1} + \delta_t,\ t = 2,...,n \tag{2.34}$$

$\delta_t$ 满足 $\mathrm{E}\left(\delta_t\right) = 0$； $\mathrm{var}\left(\delta_t\right) = \tau^2,\ t = 2,...,n$； $\mathrm{cov}\left(\delta_i, \delta_j\right) = 0,\ i \neq j$ 。

由此构造迭代模型函数为：

$$lpm_t-\rho lpm_{t-1}=\theta_1\times\left(\exp\left(-\theta_2/trg_t\right)-\rho\exp\left(-\theta_2/trg_{t-1}\right)\right)+\theta_3\times\left(w_t-\rho w_{t-1}\right)$$
$$+\theta_4\times\left(t_t-\rho t_{t-1}\right)+\theta_5\times\left(pc_t-\rho pc_{t-1}\right)+\theta_6\times\left(ep_t-\rho ep_{t-1}\right)+\theta_7\times\left(id_t-\rho id_{t-1}\right) \quad (2.35)$$

先取$\rho$为.417，当作模型(2.35)常数建立非线性回归模型，对应迭代模型 1。再取$\rho$初值为.417，当作模型(2.35)参数建立非线性回归模型，对应迭代模型 3。结合前两个模型结果，取$\rho$为.5，当作模型(2.35)常数建立非线性回归模型，对应迭代模型 2。

表 2.10 中标记出当显著水平$\alpha$=.05 时与标准化残差绝对值显著相关的部分变量。当显著水平$\alpha$=.01 时标准化残差绝对值与$trg$变量无相关性。

$\rho$过大时迭代模型异方差性显著，$\rho$过小时迭代模型自相关性显著。其中，迭代模型 2 最佳。

**表 2.10 迭代模型相关性检验表**

| 迭代模型 | $\rho$ | $\hat{y}$ | $trg$ | $w$ | $pc$ | $id$ | $cor\left(e_t,e_{t-1}\right)$ |
|---|---|---|---|---|---|---|---|
| 1 | .417 | - | -.089 | .141 | .154 | -.093 | .141 |
| 2 | .5 | - | -.093 | .155 | .164 | -.099 | - |
| 3 | .638 | .138 | -.106 | .154 | .172 | - | - |

确定迭代模型 2 为迭代模型。得到迭代模型为：

$$lpm_t-0.5lpm_{t-1}=\theta_1\times\left(\exp\left(-\theta_2/trg_t\right)-0.5\exp\left(-\theta_2/trg_{t-1}\right)\right)+\theta_3\times\left(w_t-0.5w_{t-1}\right)$$
$$+\theta_4\times\left(t_t-0.5t_{t-1}\right)+\theta_5\times\left(pc_t-0.5pc_{t-1}\right)+\theta_6\times\left(ep_t-0.5ep_{t-1}\right)+\theta_7\times\left(id_t-0.5id_{t-1}\right) \quad (2.36)$$

根据模型(2.36)建立非线性回归模型。迭代 3 步后确认迭代收敛。迭代模型固有曲率与参数效应曲率均小于三种临界值，认为曲率近似为 0，确认该模型非线性强度很低。所有参数估计的 Box 偏差很小，Box 百分偏差均小于 0.5%，认为参数估计较为合理。

**表 2.11 迭代模型迭代信息表**

| 步数 | $\hat{\theta}_1$ | $\hat{\theta}_2$ | $\hat{\theta}_3$ | $\hat{\theta}_4$ | $\hat{\theta}_5$ | $\hat{\theta}_6$ | $\hat{\theta}_7$ | 残差平方和 |
|---|---|---|---|---|---|---|---|---|
| 3 | 45.763 | 0.348 | -0.021 | -0.003 | -0.009 | -0.059 | 7.198 | 4504.411 |

表 2.12 迭代模型非线性强度度量

| 曲率 | 曲率类型 | 曲率值 | 临界值 | 0.5×临界值 | 0.2×临界值 |
|---|---|---|---|---|---|
| $\rho K_{MS}^{N}$ | 固有曲率 | 0.004 | 0.702 | 0.351 | 0.140 |
| $\rho K_{MS}^{P}$ | 参数效应曲率 | 0.013 | 0.702 | 0.351 | 0.140 |

表 2.13 迭代模型参数估计的 **Box** 偏差度量

| 参数估计 | $\hat{\theta}_1$ | $\hat{\theta}_2$ | $\hat{\theta}_3$ | $\hat{\theta}_4$ | $\hat{\theta}_5$ | $\hat{\theta}_6$ | $\hat{\theta}_7$ |
|---|---|---|---|---|---|---|---|
| Box 偏差 | 0.001 | 0.001 | 0.00001 | 0.000 | 0.000 | -0.00002 | -0.0001 |
| Box 百分偏差 | 0.002% | 0.302% | -0.042% | -0.039% | 0.003% | 0.032% | -0.002% |

进行残差分析。在图 2.9 中，所有标准化残差均在(-3,3)范围内，没有异常点。残差均值为 $1.57\times10^{-5}$，认为残差满足零均值。在 Spearman 相关性检验中，当显著水平 $\alpha=.01$ 时，标准化残差绝对值与 $\hat{y}$、$t$、$ep$、$trg$、$id$ 等无显著相关性。而变量 $w$、$pc$ 与标准化残差绝对值之间有显著的相关性，但是相关系数较小，说明标准化残差绝对值与变量 $w$、$pc$ 之间的相关性显著但不强。在标准化残差与变量 $w$、$pc$ 图 2.13 中，可以看出存在一定的异方差性，但异方差性的强度不大。故认为迭代模型存在显著但轻微的异方差性。

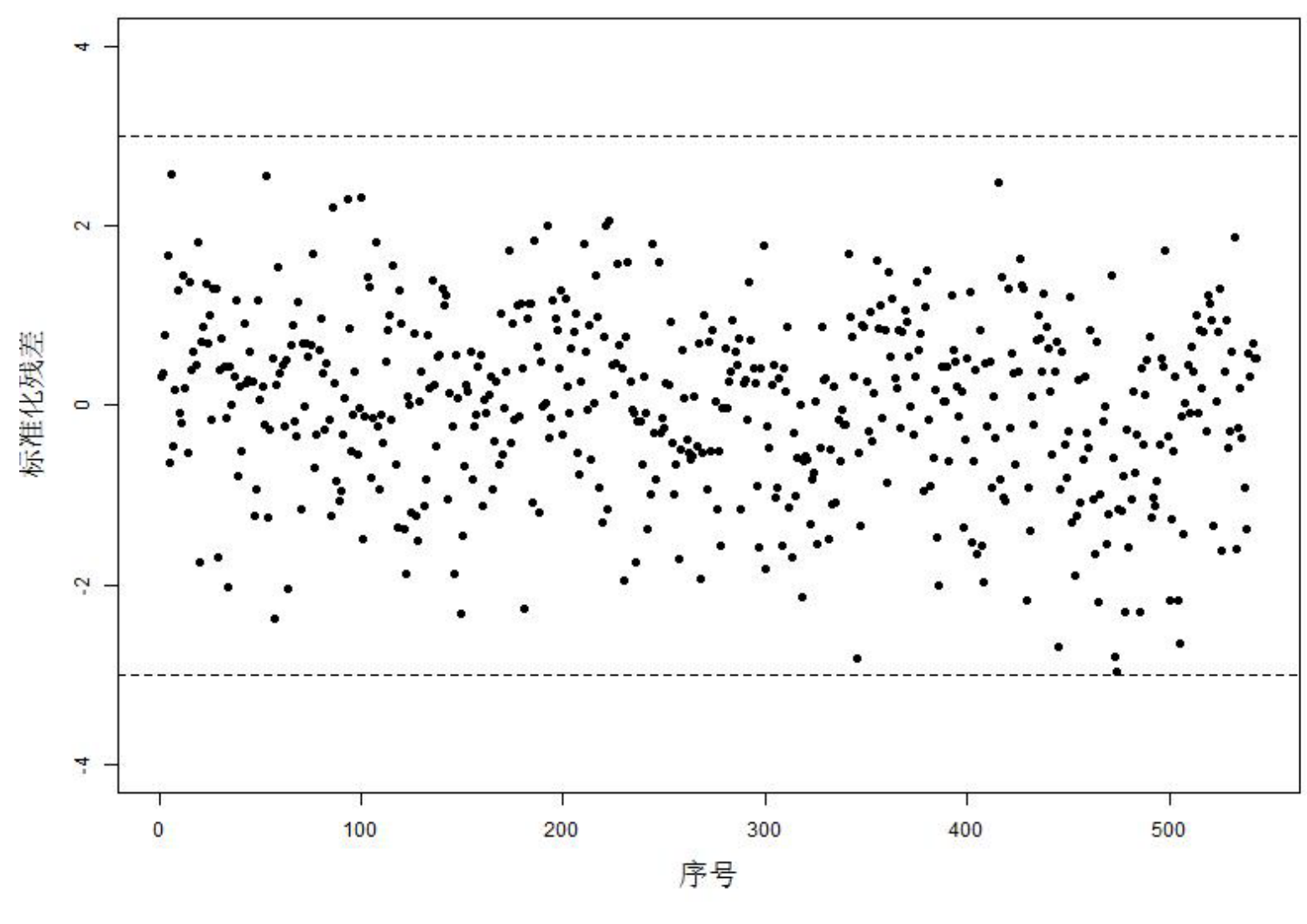

图 2.9 标准化残差图

表 2.14 标准化残差绝对值与拟合值、各自变量的 **Spearman** 相关系数表

| 相关系数 | $\hat{y}$ | $t$ | $trg$ | $w$ | $pc$ | $ep$ | $id$ |
|---|---|---|---|---|---|---|---|
| 初始模型 | -.142 | -.007 | -.156 | .154 | .168 | -.017 | - |
| $id$ 模型 | -.000 | .046 | -.067 | .019 | .098 | -.029 | -.049 |
| 迭代模型 | -.000 | .047 | -.093 | .155 | .164 | .040 | -.099 |
| $P-$值 | .999 | .274 | .030 | .0003 | .0001 | .351 | .022 |

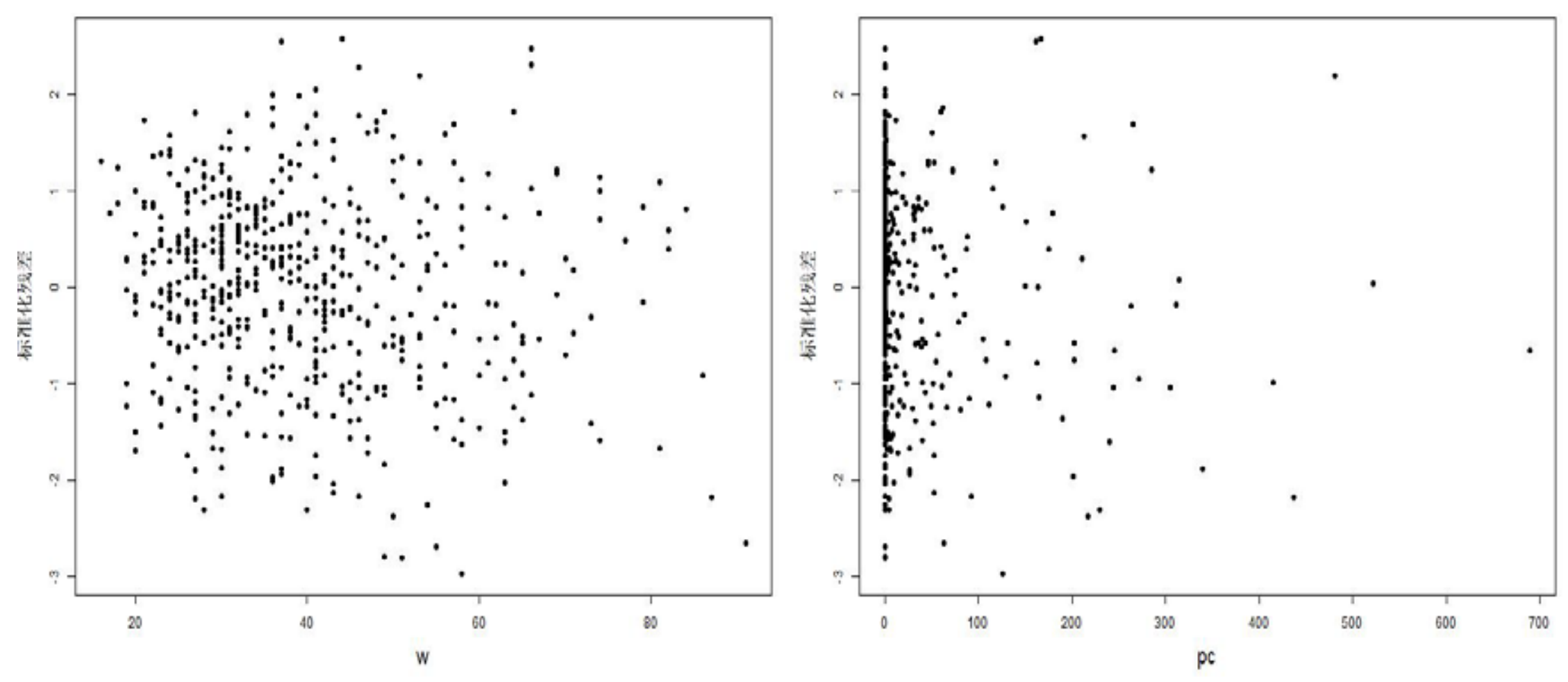

图 2.10 标准化残差与 $w$、 $pc$ 的走势图

在时序前一个标准化残差 $e_t$ 与时序后一个标准化残差 $e_{t-1}$ 图 2.11 中，标准化残差不存在显著自相关性。对时序前后相邻的两组标准化残差做 Pearson 相关性检验，相关系数为. 078，$P-$值为. 068。当显著水平 $\alpha=.01$ 时认为标准化残差不存在自相关性。

虽然迭代模型的异方差性比 $id$ 模型强，但迭代模型的自相关性基本被消除。因为 $e_t$ 的性质会影响 $\delta_t$ 的性质，所以迭代模型中的异方差性与自相关性很难同时消除，只能寻找一个较为均衡的状态使得模型关于两性质能基本满足基本假设。

综上所述，迭代模型存在轻微异方差性，无自相关性。与初始模型相比，迭代模型有效降低了异方差性，基本消除了自相关性，因此可以认为迭代模型基本符合非线性回归模型关于随机误差项零均值、等方差、不相关的基本假设。

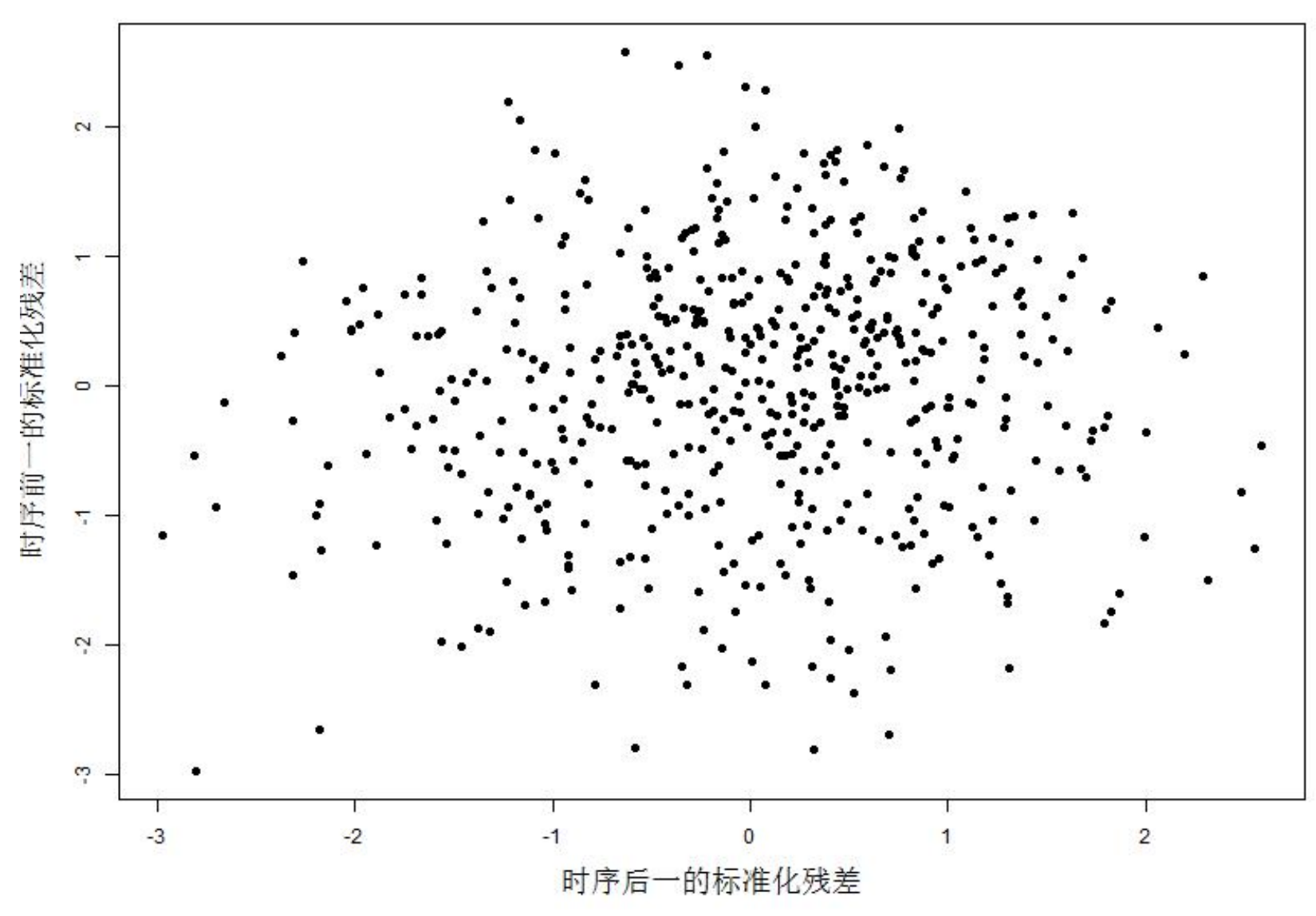

图 2.11 时序前后相邻的标准化残差图

对残差做 Kolmogorov-Smirnov 正态性检验，$D$=0.246，$P-$值$<2.2\times10^{-16}$，故当显著水平$\alpha$=.05 时残差不服从正态分布，迭代模型不满足正态假设。

### 2.2.5 模拟研究

由于迭代模型不满足正态假设，故采用模拟的方法研究参数估计的优良性。鉴于残差分布未知，制定模拟研究步骤为：

(1)以变量$id$为指标在样本资料阵中分层抽样，抽样样本量固定为 450。

(2)在一次模拟中，根据迭代模型求解最小二乘估计并计算模拟残差。在满足模拟残差与迭代模型中对应位置的样本残差完全不相同的前提下，判断模拟残差与样本残差是否来自同一分布。

(3)重复步骤(2) 1000 次，得到 1000 个各参数估计的模拟样本估计。若 1000 次模拟中接受模拟残差与样本残差来自同一分布的频率足够大，则接受 1000 次模拟结果，否则重复步骤(2)。

(4)根据 1000 次模拟样本信息研究样本参数估计的优良性。

预模拟 100 次，判断迭代模型是否收敛以及其非线性强度是否足够低。在 100 次预模拟中模型全部达到收敛状态，且非线性强度很低。见表 2.15 与表 2.16。

**表 2.15 预模拟的迭代信息表**

| 重复次数 | 收敛次数 | 收敛率 | 平均迭代步数 | 最大迭代步数 | 最小迭代步数 |
|---|---|---|---|---|---|
| 100 | 100 | 100% | 2.98 | 3 | 2 |

**表 2.16 预模拟的曲率大小比较信息表**

| 曲率 | 次数 | 通过率 | 曲率 | 次数 | 通过率 |
|---|---|---|---|---|---|
| $\rho K_{MS}^{N}<$临界值 | 100 | 100% | $\rho K_{MS}^{P}<$临界值 | 100 | 100% |

在一次模拟中，确认迭代模型收敛并且非线性强度很低。对比模拟残差与对应位置的样本残差，两者完全不相同。在 Kolmogorov-Smirnov 双样本检验中，当显著水平 $\alpha$=.05 时，$D$=0.033，$P-$值=.95，接受模拟残差与样本残差来自同一分布的假设。在图 2.12 中，模拟残差与样本残差的经验分布非常接近，故接受一次模拟结果。

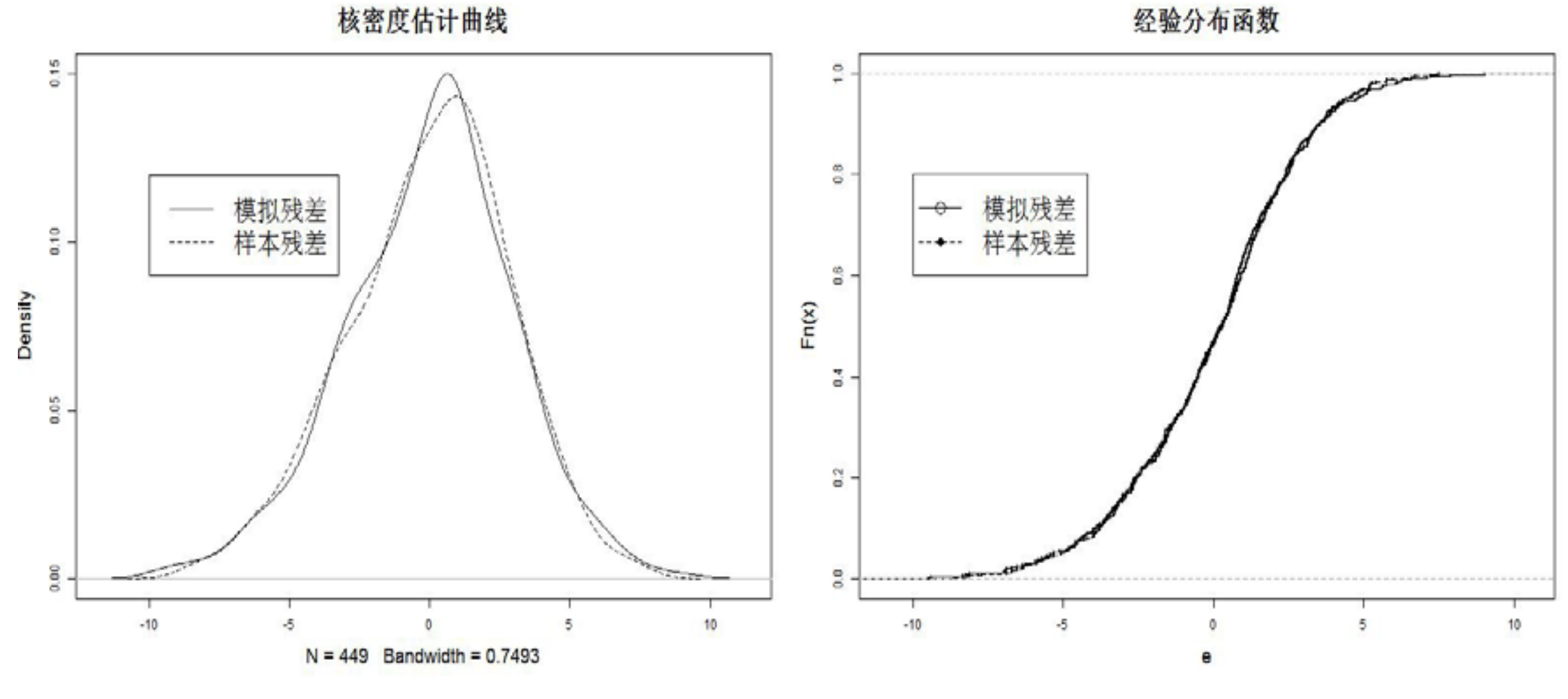

图 2.12 判断总体残差与样本残差是否来自同一分布

重复模拟 1000 次。1000 次迭代全部收敛，曲率小于临界值的通过率为 100%，模拟残差与对应位置的样本残差完全不相同。在 Kolmogorov-Smirnov 双样本检验中，1000 次模拟的 $P-$值均同时大于.05 与.5，接受模拟残差与样本残差来自同一分布的假设。

从表 2.17 可知，所有模拟参数估计都是有偏的，但偏差数量级很小。同时所有模拟参数估计的均方误差很小，接近于 0。综上所述，所有样本参数估计是有偏的、有效的。对于迭代模型而言，所有自变量对于因变量的影响是有效的。

模拟研究的偏差与 Box 偏差不同是因为 Box 偏差只是近似偏差；但对于大部分变量偏差，两者数量级接近，说明 Box 偏差在估计近似偏差中仍是有效的。

**表 2.17 模拟参数估计的优良性检验**

| | $\hat{\theta}_1*$ | $\hat{\theta}_2*$ | $\hat{\theta}_3*$ | $\hat{\theta}_4*$ | $\hat{\theta}_5*$ | $\hat{\theta}_6*$ | $\hat{\theta}_7*$ |
|---|---|---|---|---|---|---|---|
| 偏差 | -0.090 | -0.003 | -0.002 | 0.002 | 0.00003 | -0.001 | 0.172 |
| 标准差 | 0.276 | 0.145 | 0.006 | 0.002 | 0.001 | 0.011 | 0.175 |
| 均方误差 | 0.084 | 0.021 | 0 | 0 | 0 | 0 | 0.060 |

对于非线性回归模型，有偏参数可能会帮助模型产生更好的拟合效果。但鉴于迭代模型的非线性强度很低，模型性态接近线性性态，因此有偏参数在非线性回归模型进行预测时可能会产生偏差较大的预测期望曲面，故考虑修偏参数估计。见表 2.18。从表 2.19 可知，修偏参数估计仍是有效的，故认为迭代模型的修偏参数估计是优良的。

**表 2.18 修偏参数估计**

| 参数估计 | $\hat{\theta}_1$ | $\hat{\theta}_2$ | $\hat{\theta}_3$ | $\hat{\theta}_4$ | $\hat{\theta}_5$ | $\hat{\theta}_6$ | $\hat{\theta}_7$ |
|---|---|---|---|---|---|---|---|
| 修偏前 | 45.763 | 0.348 | -0.021 | -0.003 | -0.00915 | -0.059 | 7.198 |
| 修偏后 | 45.672 | 0.344 | -0.023 | -0.001 | -0.00912 | -0.060 | 7.370 |

**表 2.19 修偏后模拟参数估计的有效性检验**

| | $\hat{\theta}_1*$ | $\hat{\theta}_2*$ | $\hat{\theta}_3*$ | $\hat{\theta}_4*$ | $\hat{\theta}_5*$ | $\hat{\theta}_6*$ | $\hat{\theta}_7*$ |
|---|---|---|---|---|---|---|---|
| 标准差 | 0.276 | 0.145 | 0.006 | 0.002 | 0.001 | 0.011 | 0.175 |
| 均方误差 | 0.076 | 0.021 | 0 | 0 | 0 | 0 | 0.031 |

将修偏参数估计回代入原迭代模型中，计算修偏迭代模型的非线性曲率。从表 2.20 可知，修偏迭代模型的非线性强度仍然很低。

**表 2.20 修偏迭代模型非线性强度度量**

| 曲率 | 曲率类型 | 曲率值 | 临界值 | 0.5×临界值 | 0.2×临界值 |
|---|---|---|---|---|---|
| $\rho K_{MS}^{N}$ | 固有曲率 | 0.004 | 0.702 | 0.351 | 0.140 |
| $\rho K_{MS}^{P}$ | 参数效应曲率 | 0.013 | 0.702 | 0.351 | 0.140 |

进行残差分析。所有标准化残差均在(-3,3)范围内，没有异常点。残差均值为 0.008，认为残差满足零均值。在 Spearman 相关性检验中，当显著水平 $\alpha$=.01 时，标准化残差绝对值与 $\hat{y}$、$t$、$ep$、$trg$、$id$ 等无显著相关性。而变量 $w$、$pc$ 与标准化残差绝对值之间有显著的相关性，但是相关系数较小，说明标准化残差绝对值与变量 $w$、$pc$ 之间的相关性不强。这与原迭代模型的 Spearman 相关性检验结果相似，故认为修偏迭代模型存在显著但轻微的异方差性。

**表 2.21 标准化残差绝对值与拟合值、各自变量的 Spearman 相关系数表**

| 变量 | $\hat{y}$ | $t$ | $trg$ | $w$ | $pc$ | $ep$ | $id$ |
|---|---|---|---|---|---|---|---|
| 相关系数 | -.025 | .045 | -.095 | .151 | .161 | .037 | -.084 |
| $P-$值 | .559 | .293 | .028 | .0004 | .0002 | .387 | .051 |

对时序前后相邻的两组标准化残差做 Pearson 相关性检验，相关系数为.067，$P-$值为.120。当显著水平 $\alpha$=.01 时认为标准化残差不存在自相关性。

综上所述，修偏迭代模型基本符合随机误差项零均值、等方差、不相关的基本假设。修偏没有导致迭代模型产生异常性质，因此修偏是可行的。

修偏后，对数 $PM_{2.5}$ 拟合值与对数 $PM_{2.5}$ 观测值之差的平方和为 6784.372，回归平方和为 7102.168；修偏前，对数 $PM_{2.5}$ 拟合值与对数 $PM_{2.5}$ 观测值之差的平方和为 6832.748，回归平方和为 6864.762。修偏减小了残差平方和并增大了回归平方和，因此修偏是有效的。

## 2．2．6 拟合分析

根据修偏参数估计计算修偏拟合值。在迭代模型修偏拟合值对观测值的拟合图 2.13 中，迭代模型的表现较为稳定，模型形态与样本点的分布相近，模型能够有效刻画因变量的真实变化趋势；当样本点在均值附近时模型拟合效果较好，而在高浓度处模型会出现低估现象，在低浓度处时模型会出现高估现象。

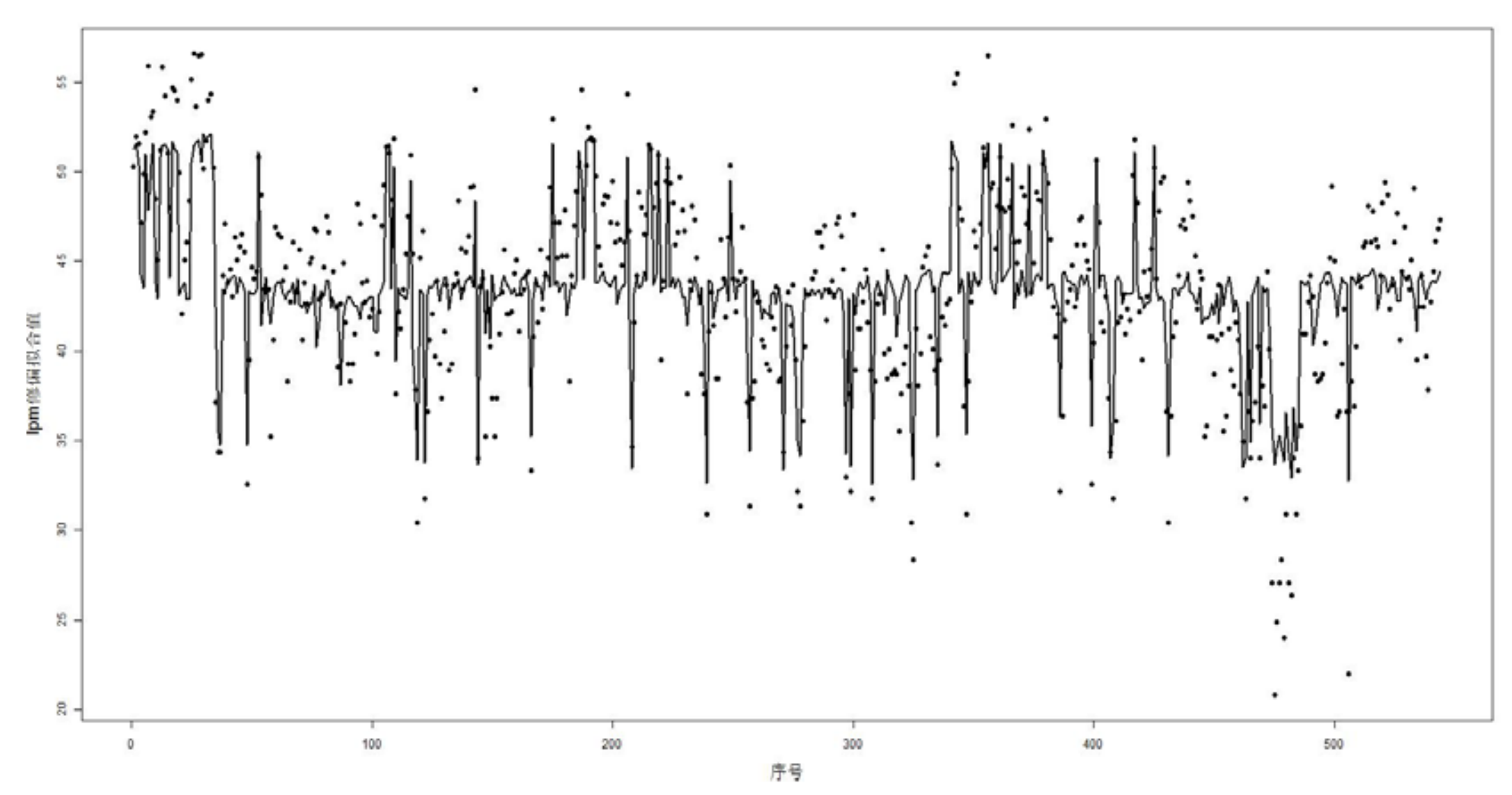

图 2.13 2014 年-2016 年对数 $PM_{2.5}$ 修偏拟合曲线对样本点的拟合图

记修偏拟合值与观测值之差的绝对值为修偏残差绝对值。在修偏残差绝对值的直方图 2.14 中，50%的修偏残差绝对值小于 3，90%的修偏残差绝对值小于 6，在绝大部分样本点处修偏拟合值与观测值之差的绝对值小于 5，说明修偏拟合值与观测值之间的差异较小，认为修偏拟合值对观测样本的拟合效果较好。

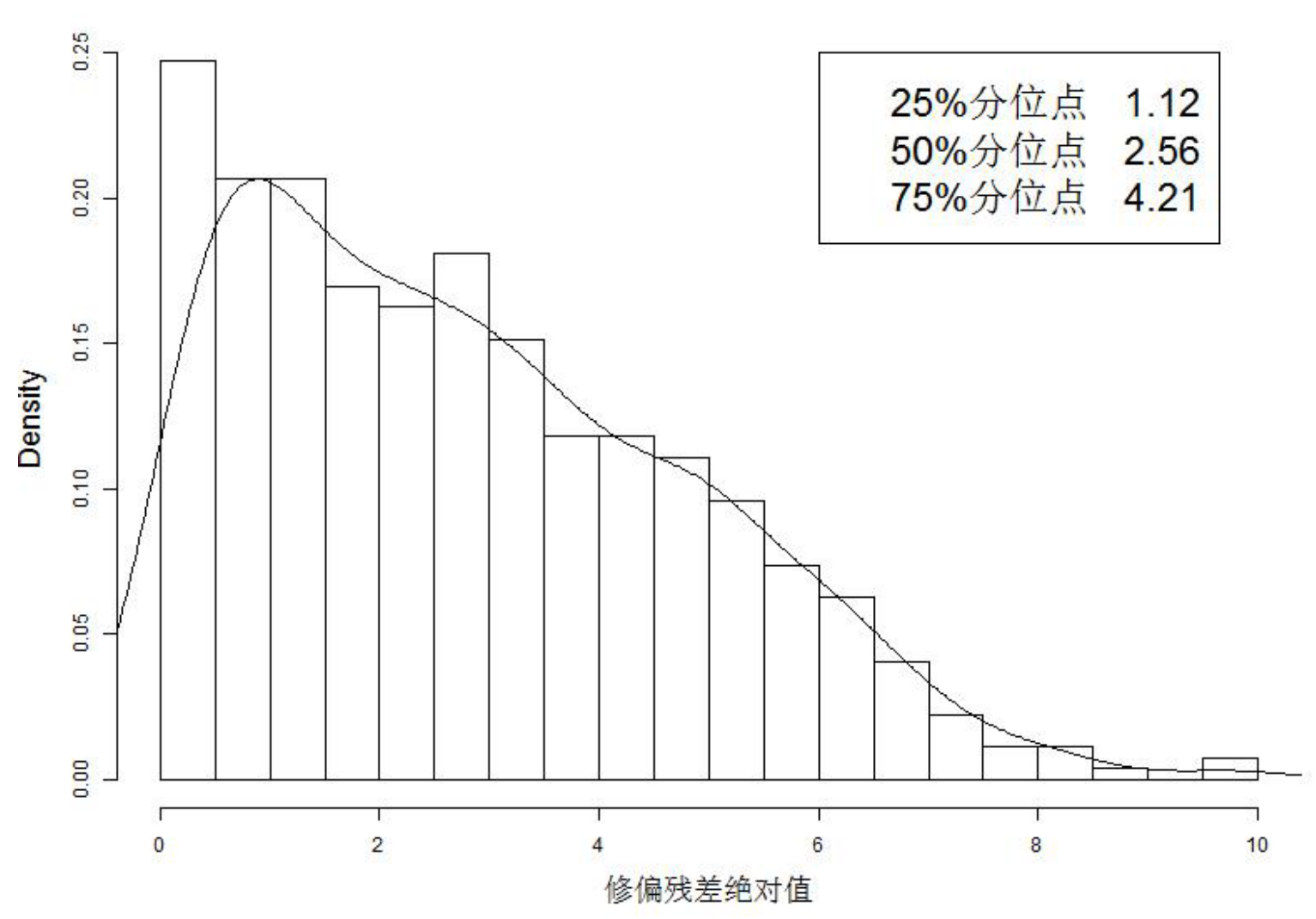

图 2.14 2014 年-2016 年修偏残差绝对值的直方图

综上所述，迭代模型的修偏参数估计是优良的，修偏拟合效果是显著的。此时迭代模型为：

$$
\begin{aligned}
\widehat{lpm} = 45.67223\exp\left(-0.34431/trg\right) - 0.02258w - 0.00109t \\
-0.00912pc - 0.05976ep + 7.36975id
\end{aligned},
$$

$$
id = \begin{cases} -1, & lpm \le 35 \\ 0, & 35 < lpm \le 50 \\ 1, & lpm > 50 \end{cases} \tag{2.37}
$$

将模型(2.37)还原，得到$PM_{2.5}$浓度的非线性回归单值预报模型为：

$$
\widehat{pm} = \exp\begin{pmatrix} 4.567223\exp\left(-0.34431/trg\right) - 0.002258w - 0.000109t \\ -0.000912pc - 0.005976ep + 0.736975id \end{pmatrix},
$$

$$
id = \begin{cases} -1, & pm \le e^{3.5} \\ 0, & e^{3.5} < pm \le e^{5} \\ 1, & pm > e^{5} \end{cases} \tag{2.38}
$$

简称$PM_{2.5}$单值预报模型。

在$PM_{2.5}$单值预报模型对样本的拟合图 2.15 中，单值预报模型的整体形态较为稳定且与样本曲线形态相吻合，基本能准确体现$PM_{2.5}$浓度的真实变化趋势。在部分高浓度处真实模型会出现严重低估现象，可能是因为存在极端天气。同时在部分低浓度处模型也存在轻度的高估现象。因此$PM_{2.5}$单值预报模型对中等浓度天气的拟合效果更好，而对高低浓度天气的预报准确度有限。

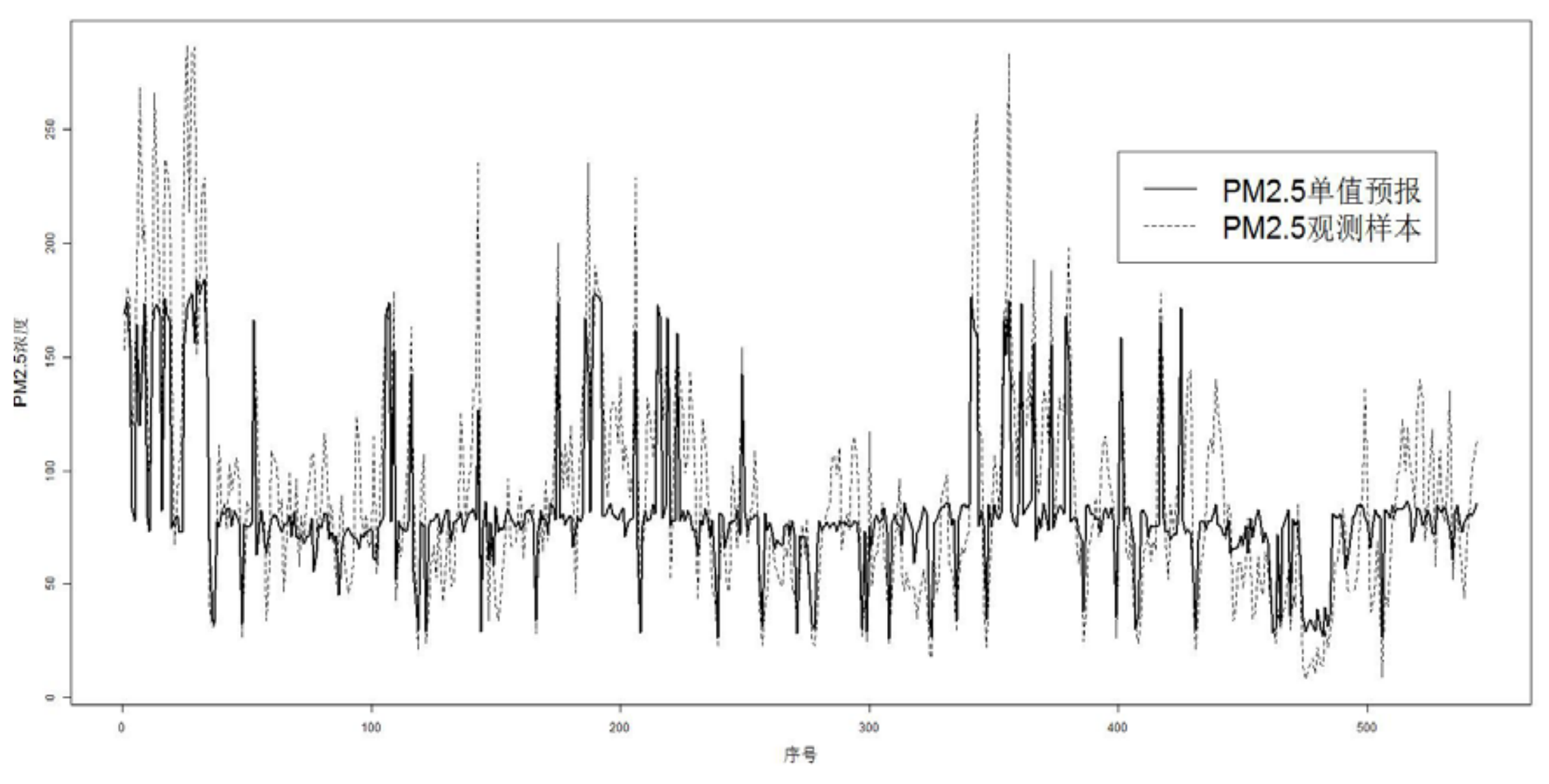

图 2.15 2014 年-2016 年$PM_{2.5}$单值预报模型对观测样本的拟合图

## 2.3 区间预报模型

由于$PM_{2.5}$单值预报模型的非线性强度很低，模型性态近似线性性态，因此可以采用线性区域作为模型的近似区间估计。鉴于单值预报模型擅长预报中等浓度同时具有高估低浓度和低估高浓度的特性，为发挥单值预报模型的最大效用，故采用分段区间建立$PM_{2.5}$区间预报模型。

根据国家环境空气质量标准[13]，$PM_{2.5}$的 24 小时平均浓度二级浓度限值为 75 $\mu g/m^3$，一级浓度限值为 35 $\mu g/m^3$。仅考虑$PM_{2.5}$浓度对空气质量的影响。当$PM_{2.5}$浓度在一级浓度限值以下时，空气质量为优，对人体危害极小；当$PM_{2.5}$浓度在二级浓度限值以上时，存在空气污染现象，尤其当$PM_{2.5}$浓度达到 150 $\mu g/m^3$ 以上时，空气污染较为严重，属于$PM_{2.5}$高污染天气，对人体危害较大。

结合实际背景与$PM_{2.5}$单值预报模型，定义$PM_{2.5}$区间预报模型 $I$ 为：

$$pm \in I = \begin{cases} (0,\ 35), & \widehat{pm} < 35 \\ \left[\widehat{pm} - r, \widehat{pm} + 1.5r\right], & 35 <= \widehat{pm} < 150, \quad r \geq 0 \\ > 150, & \widehat{pm} > 150 \end{cases} \tag{2.39}$$

当模型预报值 $\widehat{pm}$ 小于 35 $\mu g/m^3$ 时，预测真实$PM_{2.5}$浓度介于 0 ~ 35 $\mu g/m^3$，属低浓度天气，空气质量为优。当模型预报值 $\widehat{pm}$ 大于 150 $\mu g/m^3$ 时，预测真实$PM_{2.5}$浓度大于 150 $\mu g/m^3$ 且可能出现浓度超过 200 $\mu g/m^3$ 的极端天气，属高浓度天气并预警极高浓度天气，空气质量为差。当模型预报值 $\widehat{pm}$ 介于 35 ~ 150 $\mu g/m^3$ 时，预测真实$PM_{2.5}$浓度属于由$PM_{2.5}$单值预报模型预报值构成的区间内，属中等浓度天气，当$PM_{2.5}$预报浓度未超过 75 $\mu g/m^3$ 时空气质量为良，否则空气质量为较差。

$r$ 表示区间精度，其值越小则区间精度越高。不同的实际需求对 $r$ 的取值标准不同。为供参考，提出两类预报区间：精度预报区间与准确度预报区间。

精度预报区间取 $r$=20，记为 $I_1 = \begin{cases} (0,\ 35), & \widehat{pm} < 35 \\ \left[\widehat{pm} - 20, \widehat{pm} + 30\right], & 35 <= \widehat{pm} < 150 \\ > 150, & \widehat{pm} > 150 \end{cases}$。

准确度预报区间取 $r$=30，记为 $I_2 = \begin{cases} (0,\ 35), & \widehat{pm} < 35 \\ \left[\widehat{pm} - 30, \widehat{pm} + 45\right], & 35 <= \widehat{pm} < 150 \\ > 150, & \widehat{pm} > 150 \end{cases}$。

两类预报区间的应用需结合$PM_{2.5}$单值预报模型。定义样本包含率为属于预报区间的$PM_{2.5}$观测样本个数与$PM_{2.5}$观测样本总数的百分比。对于 2014 年至 2016 年全三年建模数据集，$I_1$的样本包含率能达到 60%以上，$I_2$的样本包含率能达到 80%以上。见表 2.22。由此确定精度预报区间的提议标准为 60%，准确度预报区间的提议标准为 80%。

表 2.22 两类预报区间的样本包含率

| 预报区间 | 2014 | 2015 | 2016 | 全三年 |
|---|---|---|---|---|
| $I_1$ | 74.0% | 61.9% | 62.6% | 66.2% |
| $I_2$ | 88.4% | 82.9% | 78.6% | 83.3% |

两类预报区间的实际预报效果图 2.16 可以形象展现两类预报区间的实际预报差异。准确度区间对边界样本点的覆盖效果更好，但也扩大了其他样本点的实际区间长度;而精度区间的区间长度相对较为合理，但易漏掉边界样本点。

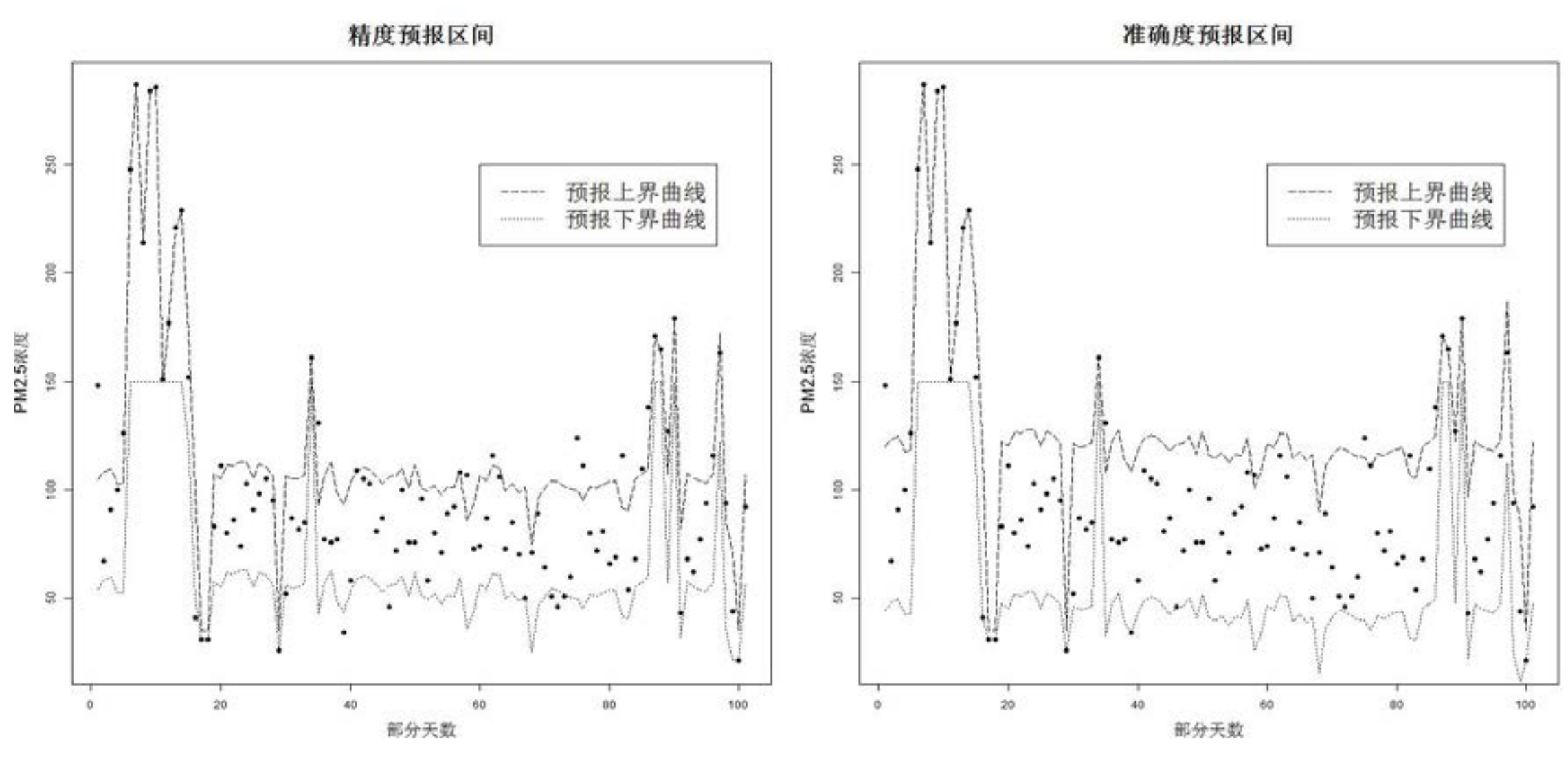

图 2.16 两类预报区间在连续部分天数内的预报效果图

## 2.4 模型检验

模型检验以验模数据集展开，检验$PM_{2.5}$浓度预报模型的预报效果，包括检验

$PM_{2.5}$单值预报模型的独立预报效果以及检验$PM_{2.5}$区间预报模型的独立预报效果，即检验两类预报区间是否分别达到 60%与 80%的提议标准。

在对 2017 年检验样本的拟合图 2.17 中，$PM_{2.5}$单值预报模型能较好拟合检验样本的变化趋势，模型趋势基本吻合样本点分布形态，但在某些样本点处模型拟合偏差较大。整体来看，单值预报模型能达到预期拟合效果。$PM_{2.5}$单值预报模型的优点是模型曲线分布在均值附近，整体上与真实变化趋势吻合较好且能有效预报中等浓度$PM_{2.5}$天气。其缺点是对高值与低值的预测准确度尚待提升，开展独立预报时可能出现较大误差。

检验$PM_{2.5}$区间预报模型的预报效果。对于 2017 年验模数据集，$I_1$的样本包含率超过 60%，$I_2$的样本包含率超过 80%，均达到提议标准。综上所述，认为$PM_{2.5}$区间预报模型通过一次检验。见表 2.23。

表 2.23 **2017 年两类预报区间的样本包含率**

| 预报区间 | $I_1$ | $I_2$ |
| --- | --- | --- |
| 样本包含率 | 65.4% | 83.0% |

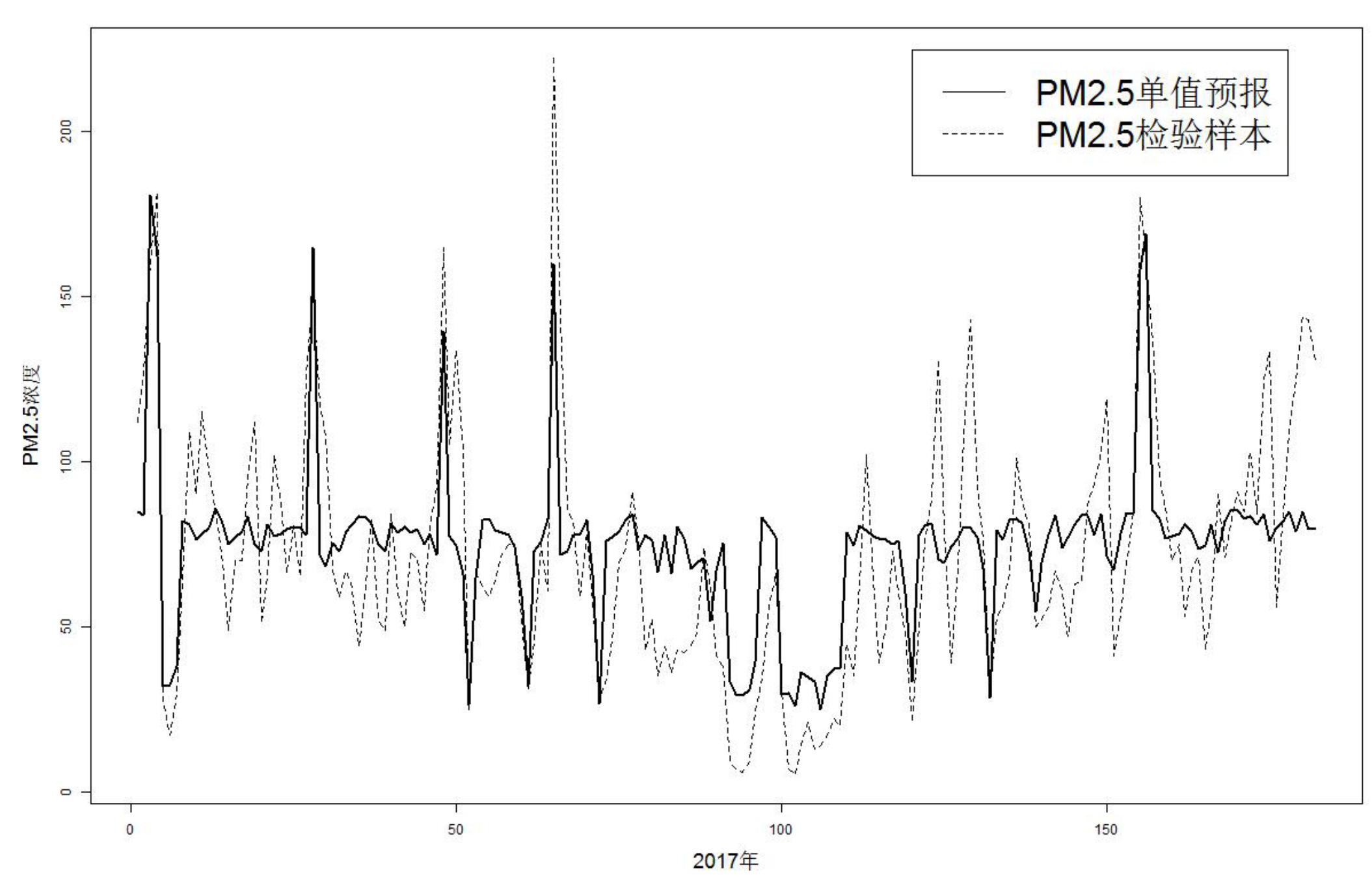

图 2.17 2017 年$PM_{2.5}$单值预报模型对检验样本的拟合图

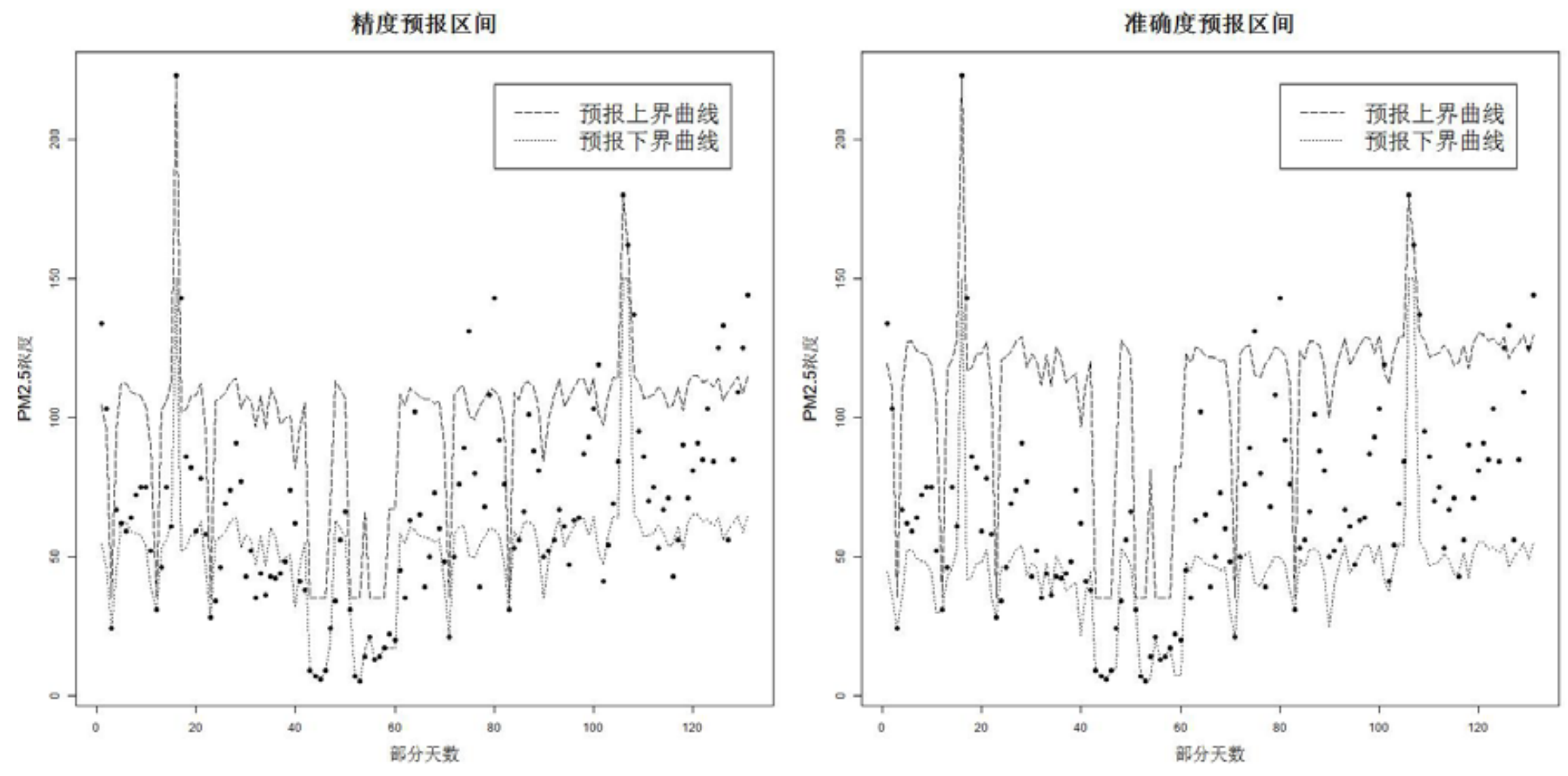

图 2.18 两类预报区间在 2017 年连续部分天数内的预报效果图

在与 $PM_{2.5}$ 单值预报模型结合后，$PM_{2.5}$ 区间预报模型有效预报高值与低值的能力得到了充分体现。同时 $PM_{2.5}$ 区间预报模型融合了 $PM_{2.5}$ 单值预报模型对于中等浓度 $PM_{2.5}$ 天气的预报特长，进一步优化了预报效果。

虽然在单值预报中单个预报值与观测值之差很小，但由于模型预报值与观测值之差呈正负交替分布于整条样本带上，因此预报区间的长度需要扩大两倍以上才能覆盖大多数检验样本。实际观测值的取值范围常常只介于预报区间某一侧端点值与 $PM_{2.5}$ 单值预报模型的预报值之间，所以 $PM_{2.5}$ 区间预报模型不能完全体现 $PM_{2.5}$ 单值预报模型的预报精度。

# 3 $PM_{2.5}$浓度预报模型的预报应用

本章主要考察$PM_{2.5}$浓度预报模型在预报应用中的实现情况。使用的预报数据集时段为2016年10月至12月与2017年1月至3月、10月至12月。

采用NCEP CFS2气象预报系统与$PM_{2.5}$浓度预报模型结合。NCEP CFS2气象预报系统全称The National Centers for Environmental Prediction (NCEP) Climate Forecast System (CFS) Version 2。目前NCEP CFS2暂未提供蒸发量预报数据，因此蒸发量数据仍采用真实数据，而温度、最高温度、最低温度、降水量、风速等预报数据采用NCEP CFS2预报数据。该系统提供每日四次逐6小时预报值，对于温度、温差、降水量等变量采用四次逐6小时预报数据的算术平均值作为单日预报数据，对于风速变量采用四次逐6小时预报数据的最大值作为单日预报数据。

## 3.1 $PM_{2.5}$浓度预报模型的预报应用

$PM_{2.5}$浓度预报模型指$PM_{2.5}$单值预报模型与$PM_{2.5}$区间预报模型组成的整体模型。$PM_{2.5}$浓度预报模型的结构为：

$$\widehat{pm} = \exp\begin{pmatrix} 4.567223\exp\left(-0.34431/trg\right) - 0.002258w - 0.000109t \\ -0.000912pc - 0.005976ep + 0.736975id \end{pmatrix},$$

$$id = \begin{cases} -1, & pm \le e^{3.5} \\ 0, & e^{3.5} < pm \le e^{5} \\ 1, & pm > e^{5} \end{cases},$$

$$I_1 = \begin{cases} \left(0,\ 35\right), & \widehat{pm} < 35 \\ \left[\widehat{pm} - 20, \widehat{pm} + 30\right], & 35 <= \widehat{pm} < 150 \\ > 150, & \widehat{pm} > 150 \end{cases},$$

$$I_2 = \begin{cases} \left(0,\ 35\right), & \widehat{pm} < 35 \\ \left[\widehat{pm} - 30, \widehat{pm} + 45\right], & 35 <= \widehat{pm} < 150 \\ > 150, & \widehat{pm} > 150 \end{cases} \tag{3.1}$$

将$PM_{2.5}$浓度预报模型与 NCEP CFS2 结合。首先采用$PM_{2.5}$观测值求取$id$变量真值。再将 NCEP CFS2 提供的温度、温差、降水量、风速预报信息代入$PM_{2.5}$浓度预报模型中计算预报值，比较预报值与观测值，从而分别考察$PM_{2.5}$浓度预报模型在单值预报与区间预报中的应用效果。

### 3．1．1 单值预报应用

在对预报应用样本的拟合图 3.1 中，$PM_{2.5}$单值预报模型能较好拟合预报应用样本的变化趋势，但单值预报的准确度不足。结合 NCEP CFS2 后，$PM_{2.5}$单值预报模型在高峰处的表现明显加强，但会出现过高预报值。同时拟合图中存在交错出现的严重高估与低估现象。

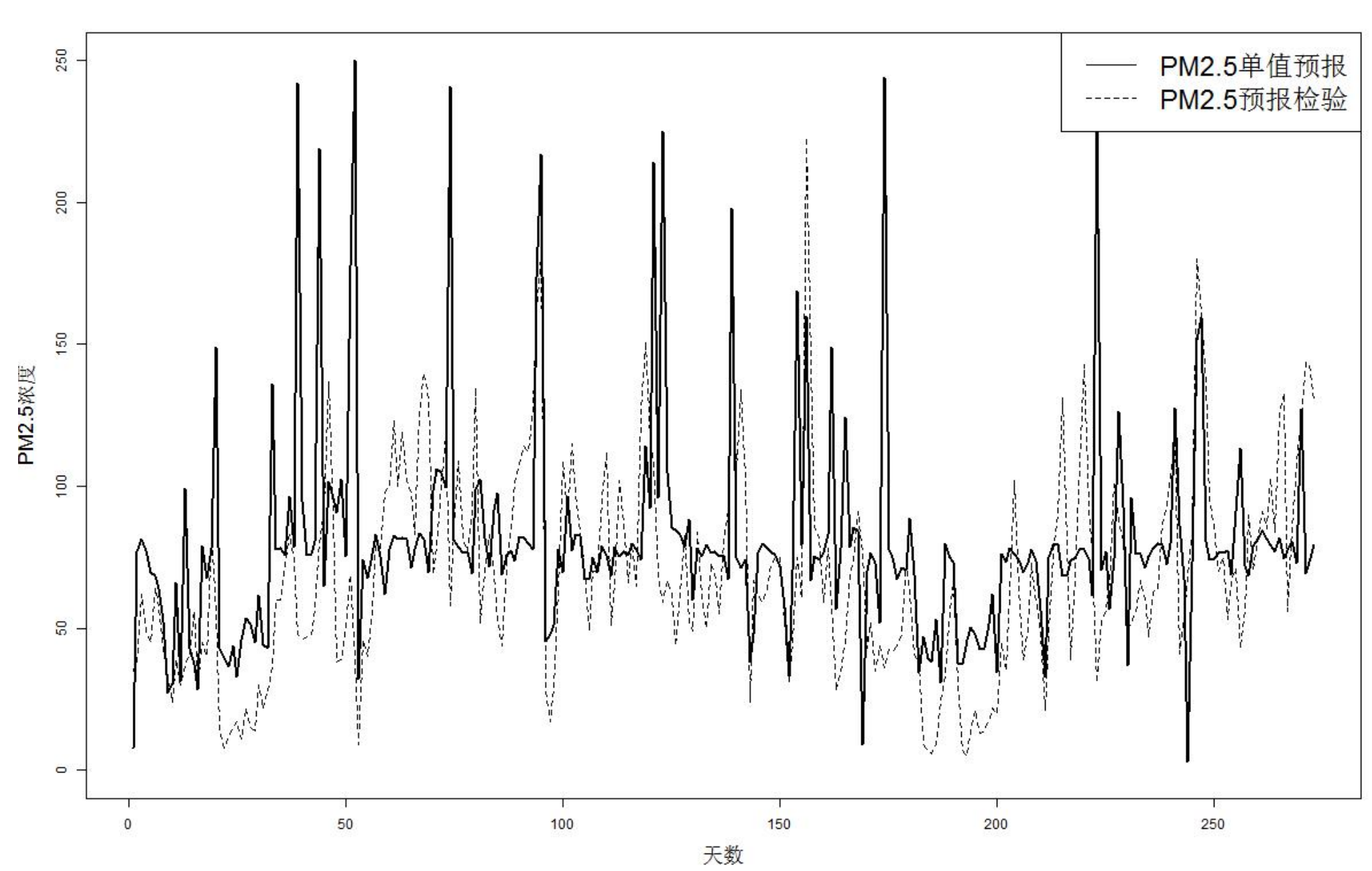

图 3.1 $PM_{2.5}$浓度预报模型对预报应用样本的拟合图

$PM_{2.5}$单值预报模型出现过高预报值很可能与模型的外推能力有关。给出各变量在建模过程与预报应用中的取值区间。见表 3.1。

其中，超出变量$t$范围的预报应用样本有 2 个，超出变量$trg$范围的预报应用样本有 107 个，超出变量$w$范围的预报应用样本有 1 个。超出变量$t$、$w$范围的预报应用样本量不大，但超出变量$trg$范围的预报应用样本量较大，故变量$trg$对预报模型的外推效果影响较大。

**表 3.1 各变量实有取值区间**

| 变量 | $t$ | $trg$ | $w$ | $pc$ | $ep$ |
| --- | --- | --- | --- | --- | --- |
| 预报应用 | $[19.3,249.6]$ | $[-20,58.5]$ | $[10.8,92.4]$ | $[0,12.28]$ | $[0,64]$ |
| 建模过程 | $[-38,243]$ | $[9,205]$ | $[16,91]$ | $[0,689]$ | $[0,64]$ |
| 单位 | 0.1℃ | 0.1℃ | 0.1m/s | 0.1mm | 0.1mm |

在预报应用中，当变量 $trg$ 取值在建模取值区间外但仍为正值时，预报值对真值的拟合效果与该变量在建模取值区间内的预报值对真值的拟合效果相近，此时预报没有出现过大偏差。

当变量 $trg$ 取值为负值时，对应 77 个样本。其中，预报模型出现 1 个大于 200、小于 300 的高预报值（208.4），对应 1 个 200 以下、100 以上的真值（182），未出现较大偏差；出现 8 个大于 300 的过高预报值，对应 7 个 100 以下的真值与 1 个大于 100、小于 200 的真值（108），出现了较大偏差，认为在这些样本点处模型外推失效；在其余样本点处，模型拟合较好。因此负值 $trg$ 不一定会完全破坏 $PM_{2.5}$ 单值预报模型的拟合效果，但对 $PM_{2.5}$ 单值预报模型的外推仍是有风险的。当负值 $trg$ 对应的 $PM_{2.5}$ 浓度预报值不超过 300 时，可以接受预报结果；但当负值 $trg$ 对应的 $PM_{2.5}$ 浓度预报值过高时则需谨慎接受预报结果，因为此时模型外推很可能失效。

变量 $trg$ 能取到负值，这与 NCEP CFS2 的预报模式、预报准确度及精度有关，说明与单值预报模型结合的气象预报系统类型会影响预报模型的实际应用效果。

鉴于单值预报模型外推失效的样本量不大，后文仍考虑模型外推的情形。

当气象预报系统能够提供更精准的预报产品时，$PM_{2.5}$ 单值预报模型的应用价值将能更充分发挥。此处结合 NCEP CFS2 说明预报应用中如何求取 $id$ 变量预报值。

在上述预报应用中，$id$ 变量预报值采用真值，但在实际预报中其真值是未知的。该变量对单值预报模型的影响显著，不能予以剔除，故需寻找较好的方法对该变量进行估计。有两种较为可行的算法:

（1） 使用前一天 $PM_{2.5}$ 浓度观测值计算 $id$ 变量预报值，即：

$$id_t=\begin{cases}-1, & pm_{t-1}\le e^{3.5}\\ 0, & e^{3.5}<pm_{t-1}\le e^{5}\\ 1, & pm_{t-1}>e^{5}\end{cases}，\ t \ \text{表示观测日；}$$

（2） 使用剔除$id$变量后的单值预报模型计算$id$变量预报值，即：

$$id=\begin{cases}-1, & \widehat{pm}'\le e^{3.5} \\ 0, & e^{3.5}<\widehat{pm}'\le e^{5} \\ 1, & \widehat{pm}'>e^{5}\end{cases}，\quad \widehat{pm}'=\exp\begin{pmatrix}4.567223\exp\left(-0.34431/trg\right)-0.002258w \\ -0.000109t-0.000912pc-0.005976ep\end{pmatrix}。$$

两种算法对预报应用样本的拟合见图3.2与图3.3。从总体拟合效果看，两种算法均能有效开展单值预报。第一种算法能更准确贴合样本变化趋势，而第二种算法的准确度稍差。

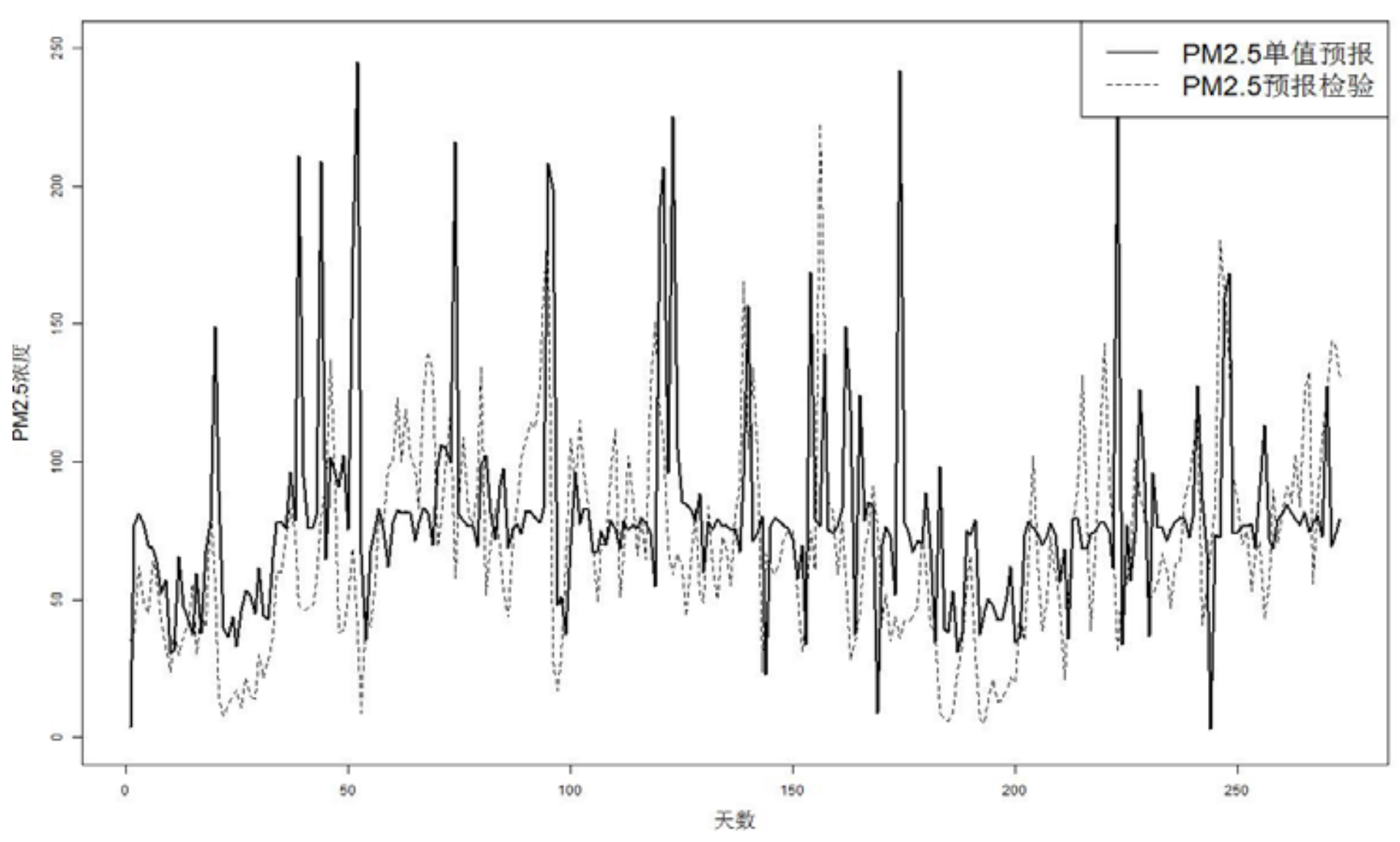

图3.2 第一种算法对预报应用样本的拟合图

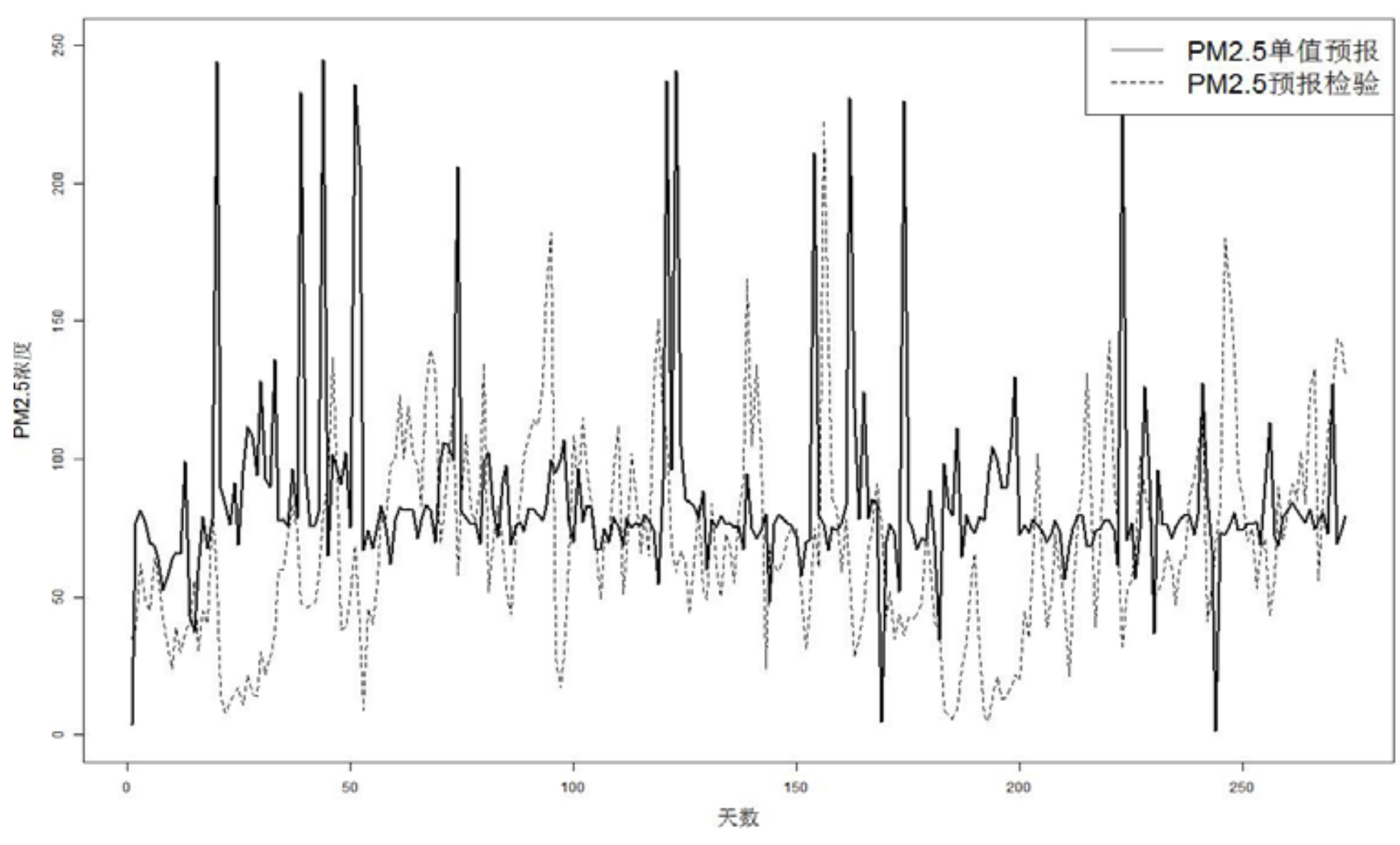

图3.3 第二种算法对预报应用样本的拟合图

进一步比较两种算法与*id*变量取真值时的预报差异。见图3.4。第一种算法与*id*变量真值拟合曲线几乎吻合，仅在部分样本点处出现微小波动，说明第一种算法能有效估计*id*变量并接近真实取值时的预报效果。第二种算法与*id*变量真值拟合曲线基本接近，也能有效估计*id*变量，但出现较大偏差的频数更高。

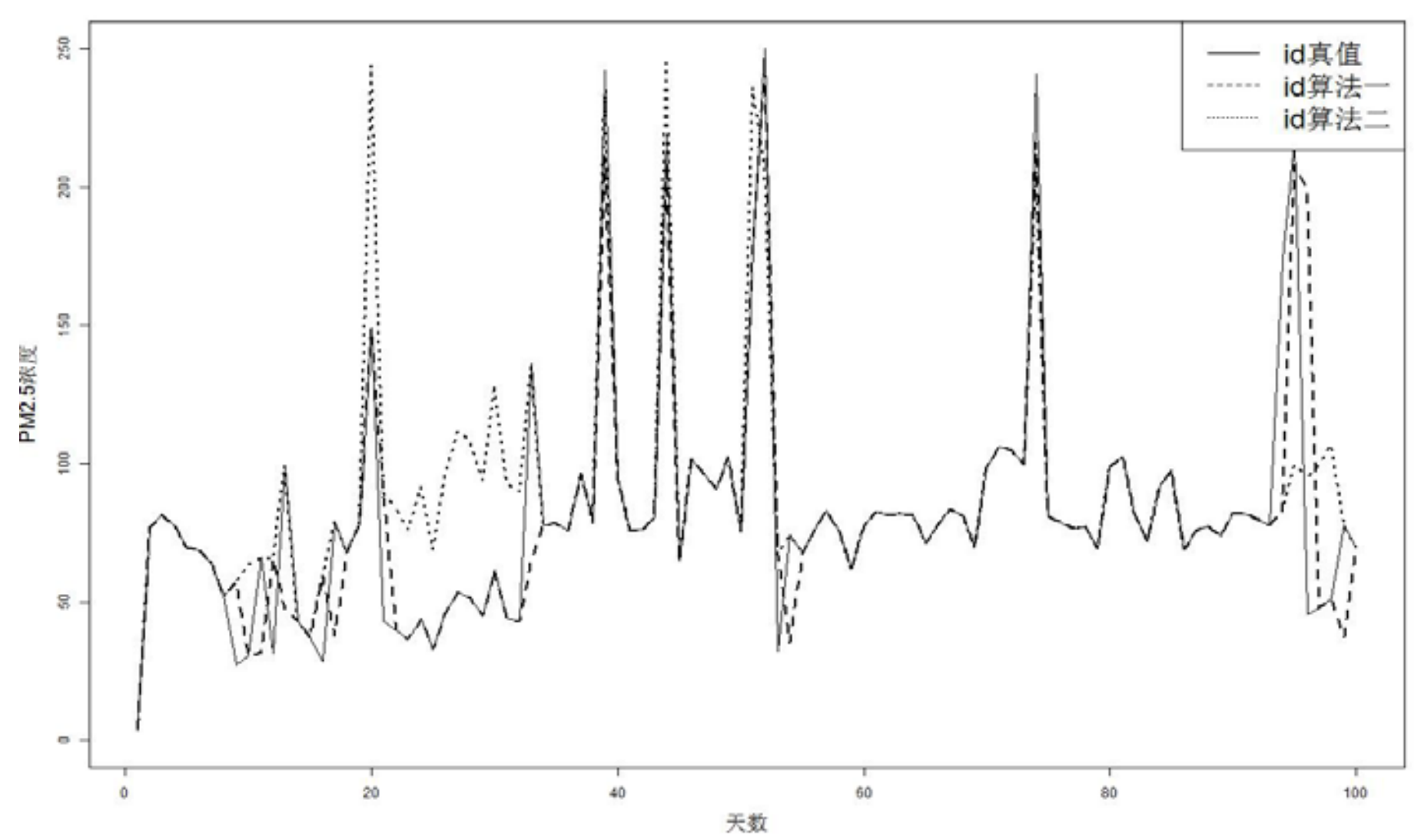

图3.4 两种算法与*id*变量真值的预报对比图

记$PM_{2.5}$预报应用样本观测值与$PM_{2.5}$单值预报模型预报值之差的绝对值为预报偏差。两种算法与*id*真值的预报偏差分位数比较见表3.2。第一种算法与*id*取真值时的预报结果相近，而第二种算法与*id*取真值时的预报结果相差较大。综上说明，对于单值预报模型的预报应用，第一种算法的预报效果更佳。

**表3.2 两种算法与*id*真值的预报偏差分位数比较**

| 分位数 | 15% | 30% | 45% | 60% | 75% | 90% |
| --- | --- | --- | --- | --- | --- | --- |
| *id*真值 | 5.87 | 12.71 | 20.49 | 27.21 | 34.67 | 58.53 |
| 算法一 | 5.91 | 11.97 | 20.64 | 27.70 | 36.41 | 65.95 |
| 算法二 | 6.80 | 14.46 | 23.62 | 31.13 | 51.92 | 84.08 |

比之模型检验，两种算法预报偏差明显增大。第二种算法的较大偏差频率更高，说明两种算法的预报能力存在显著差异。对于60%左右的预报应用样本，两种算法与*id*真值时的预报偏差不超过30。考虑到NCEP CFS2与组合预报系统的预报准确度，预报偏差在40以内是可以接受的，据此第一种算法的预报准确率能达到

75%，第二种算法的预报准确率能达到 60%。

综上所述，$PM_{2.5}$ 单值预报模型可以用于独立预报，但预报准确率尚还不足。对于 $id$ 变量预报值，建议使用第一种算法，或结合 $PM_{2.5}$ 区间预报模型的预报效果选择合适的算法。

#### 3．1．2 区间预报应用

当 $id$ 变量取真值时，对于 $PM_{2.5}$ 区间预报模型的预报应用，$I_1$ 的样本包含率能达到 50%以上，$I_2$ 的样本包含率能达到 70%以上。相比于模型检验，预报应用的样本包含率减小 10%左右，可能是因为预报应用样本量不够大，或 NCEP CFS2 的预报精度有限，或 $PM_{2.5}$ 区间预报模型与 NCEP CFS2 的结合效果不够理想。

表 3.3 区间预报应用中两类预报区间的样本包含率

| 预报区间 | $I_1$ | $I_2$ |
| --- | --- | --- |
| 样本包含率 | 52.0% | 74.4% |

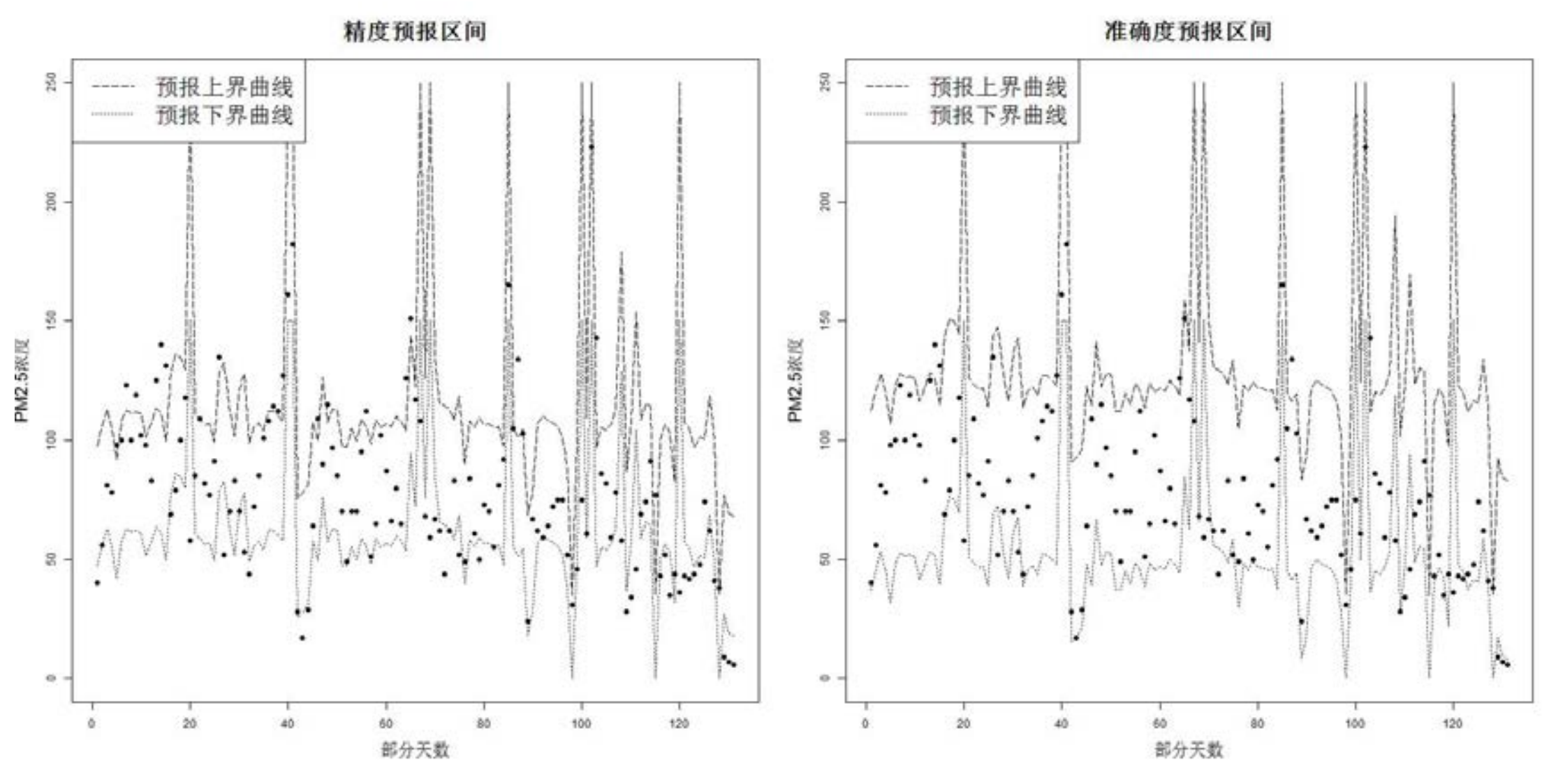

图 3.5 两类预报区间在预报检验的预报效果图

与 NCEP CFS2 结合后的区间预报模型倾向于高估真实值，因此理论预报区间应该是上半区间长度小于下半区间长度。这恰好与 $PM_{2.5}$ 区间预报模型构造方式相反，但并不表示 $PM_{2.5}$ 区间预报模型是错误的，只能说明 $PM_{2.5}$ 区间预报模型与 NCEP

CFS2 结合的整体模型没有产生预期效果。当 $PM_{2.5}$ 区间预报模型与其他气象预报系统结合时可能会产生显著优于本例和 $PM_{2.5}$ 区间预报模型的预报效果。因此在实际应用中需对与 $PM_{2.5}$ 区间预报模型结合的气象预报系统进行筛选。

再分别使用两种 *id* 变量预报值算法对区间预报应用进行分析。

采用第一种算法时区间预报应用的样本包含率相比于 *id* 取真值时小幅下降，因为前一天 $PM_{2.5}$ 浓度不一定能完全体现第二天 $PM_{2.5}$ 浓度的真实情况，前后观测值的差距将导致预报准确度下降。但样本率下降幅度不大，因此前一天 $PM_{2.5}$ 观测值在一定程度上能有效代替后一天 $PM_{2.5}$ 观测值。第一种算法在前一天 $PM_{2.5}$ 观测数据可获得的前提下便捷快速，但对真实数据有依赖性，不便于开展独立预报。

**表 3.4 第一种算法的样本包含率**

| 预报区间 | $I_1$ | $I_2$ |
| --- | --- | --- |
| 样本包含率 | 48.7% | 69.6% |

采用第二种算法时区间预报应用的样本包含率相比于 *id* 取真值时下降更多，很可能是因为 $PM_{2.5}$ 单值预报模型的预报能力还不充分，无法完全预测 $PM_{2.5}$ 真实浓度范围。虽然第二种算法在本次预报应用中的效果不如第一种算法，但应用前景更广。随着单值预报模型在未来不断提高预报效果，单值预报模型对于 $PM_{2.5}$ 真实浓度的单值与区间值的预报准确度将大幅提高，那么第二种算法的应用将会脱离对真实数据的依赖，并减少观测信息收集量，从而实现独立预报。

**表 3.5 第二种算法的样本包含率**

| 预报区间 | $I_1$ | $I_2$ |
| --- | --- | --- |
| 样本包含率 | 44.7% | 64.1% |

综合在单值预报应用与区间预报应用中两种算法的效果，同时考虑到实际应用的便捷性，可直接采用第一种算法。当单值预报模型得到更好的预报训练后，第二种算法的精度将会大大提升，届时可采用第二种算法。

## 3.2 NCEP 预报模型的预报应用

因为$PM_{2.5}$浓度预报模型与 NCEP CFS2 结合的预报效果有待加强，故根据$PM_{2.5}$浓度预报模型与 NCEP CFS2 的结合特点，提出针对 NCEP CFS2 的$PM_{2.5}$浓度预报模型，简称 NCEP 预报模型，并实现 NCEP 区间预报模型的预报应用。

建立 NCEP 预报模型为：

$$\widehat{pm} = \exp\begin{pmatrix} 4.567223\exp\left(-0.34431/trg\right) - 0.002258w - 0.000109t \\ -0.000912pc - 0.005976ep + 0.736975id \end{pmatrix},$$

$$id = \begin{cases} -1, & pm \le e^{3.5} \\ 0, & e^{3.5} < pm \le e^{5} \\ 1, & pm > e^{5} \end{cases},$$

$$I = \begin{cases} \left(0,\ 35\right), & \widehat{pm} < 35 \\ \left[\widehat{pm} - 1.5r, \widehat{pm} + r\right], & 35 <= \widehat{pm} < 150,\quad r \ge 0 \\ > 150, & \widehat{pm} > 150 \end{cases} \tag{3.2}$$

构造 NCEP 精度区间$I_1$为：

$$I_1 = \begin{cases} \left(0,\ 35\right), & \widehat{pm} < 35 \\ \left[\widehat{pm} - 30, \widehat{pm} + 20\right], & 35 <= \widehat{pm} < 150,\quad r \ge 0 \\ > 150, & \widehat{pm} > 150 \end{cases} \tag{3.3}$$

采用$id$变量真值，此时$I_1$样本包含率为 61.5%，通过精度区间检验。

构造 NCEP 准确度区间$I_2$为：

$$I_2 = \begin{cases} \left(0,\ 35\right), & \widehat{pm} < 35 \\ \left[\widehat{pm} - 45, \widehat{pm} + 30\right], & 35 <= \widehat{pm} < 150,\quad r \ge 0 \\ > 150, & \widehat{pm} > 150 \end{cases} \tag{3.4}$$

采用$id$变量真值，此时$I_2$样本包含率为 80.2%%，通过准确度区间检验。

再使用两种$id$变量预报值算法计算两类预报区间的样本包含率。见表 3.6。

NCEP 区间预报模型的样本包含率比之 NCEP CFS2 与$PM_{2.5}$区间预报模型结合时有显著提高，也更符合建模过程中的提议标准，说明 NCEP 区间预报模型的预报效果不仅优于$PM_{2.5}$区间预报模型，而且更加接近预期预报效果。同时 NCEP 区间预报

表 3.6 预报应用中两类 **NCEP** 预报区间的样本包含率

| *id* 预报值 | $I_1$ | $I_2$ |
| --- | --- | --- |
| 算法一 | 57.5% | 75.1% |
| 算法二 | 51.6% | 65.9% |
| 提议标准 | 60% | 80% |

模型的准确度与提议标准之差在扩大样本量后可能会减小，在未来可以进一步扩展 NCEP 区间预报模型的预报应用。同时证实了 $PM_{2.5}$ 预报模型与气象预报系统的结合方式会对预报效果有明显影响。

引入预报变量后 NCEP 区间预报模型的样本包含率会有一定程度的下降，因此精度区间与准确度区间的实际预报率应在 60%、70%左右；相较于提议标准，预报率将会下降 5%-10%。若想使区间预报模型达到预期预报效果，则需进一步优化区间预报模型结构并扩大样本量。对于初步预报模型而言，NCEP 区间预报模型的效果仍是显著的。

# 结　　论

## 1 结论概述

本文建立并检验了$PM_{2.5}$浓度预报模型，同时实现了该预报模型的预报应用。

$PM_{2.5}$浓度预报模型分为$PM_{2.5}$单值预报模型与$PM_{2.5}$区间预报模型。$PM_{2.5}$单值预报模型既能较为准确反映$PM_{2.5}$浓度变化趋势，又能初步实现独立预报。$PM_{2.5}$区间预报模型弥补了$PM_{2.5}$单值预报模型的不足，同时发挥单值预报特点，有效提高了预报准确度，在不同精度取值下有很大的灵活性，也能够提供空气质量的预报信息。建议使用$PM_{2.5}$区间预报模型，同时参考$PM_{2.5}$单值预报模型的预报结果，以获取较准确的$PM_{2.5}$预报浓度与对应的空气质量预报等级。

$PM_{2.5}$浓度预报模型的预报应用效果与气象预报系统类型有关。本文结合 NCEP CFS2 气象预报系统实现了$PM_{2.5}$浓度预报模型的预报应用，证明了$PM_{2.5}$浓度预报模型的可行性。为产生更好的预报效果，建议进一步对比筛选适合$PM_{2.5}$浓度预报模型的气象预报系统，或对某一气象预报系统建立特定预报机制，并不断提高预报准确度与预报精度，最终实现独立预报。

$PM_{2.5}$浓度预报模型有很好的应用前景。随着多类气象预报系统的发展，在未来可获取更多类型的气象因素预报值，从而完善$PM_{2.5}$浓度预报模型；$PM_{2.5}$浓度预报模型还可与多种气象预报系统结合，衍生出多种不同的预报机制。建议引入预报因子体现不同气象预报系统与$PM_{2.5}$浓度预报模型的组合差异，并对不同组合的预报表现进行比较。这不仅有利于开发$PM_{2.5}$浓度预报模型的潜力，而且能把握不同气象预报系统的预报特点，进而推动预报系统的发展与创新。

武汉市$PM_{2.5}$浓度非线性回归预报模型将能协助治理武汉市$PM_{2.5}$大气污染、推动提高武汉市环境空气质量，同时满足公众对$PM_{2.5}$浓度信息的需求、带来持续的社会效益。

## 2 结论要点

$PM_{2.5}$浓度预报模型，指$PM_{2.5}$单值预报模型与$PM_{2.5}$区间预报模型组成的整体预报模型。模型结构为：

$$\widehat{pm}=\exp\begin{pmatrix}4.567223\exp\left(-0.34431/trg\right)-0.002258w-0.000109t\\ -0.000912pc-0.005976ep+0.736975id\end{pmatrix},$$

$$id=\begin{cases}-1, & pm\le e^{3.5}\\ 0, & e^{3.5}<pm\le e^{5}\\ 1, & pm>e^{5}\end{cases},$$

$$I=\begin{cases}\left(0,\ 35\right), & \widehat{pm}<35\\ \left[\widehat{pm}-r,\widehat{pm}+1.5r\right], & 35<=\widehat{pm}<150, \quad r\ge 0\\ >150, & \widehat{pm}>150\end{cases}$$

$$I_1=\begin{cases}\left(0,\ 35\right), & \widehat{pm}<35\\ \left[\widehat{pm}-20,\widehat{pm}+30\right], & 35<=\widehat{pm}<150,\\ >150, & \widehat{pm}>150\end{cases}$$

$$I_2=\begin{cases}\left(0,\ 35\right), & \widehat{pm}<35\\ \left[\widehat{pm}-30,\widehat{pm}+45\right], & 35<=\widehat{pm}<150\\ >150, & \widehat{pm}>150\end{cases} \tag{1}$$

(1) $PM_{2.5}$浓度非线性回归单值预报模型，简称$PM_{2.5}$单值预报模型。模型结构为：

$$\widehat{pm}=\exp\begin{pmatrix}4.567223\exp\left(-0.34431/trg\right)-0.002258w-0.000109t\\ -0.000912pc-0.005976ep+0.736975id\end{pmatrix},$$

$$id=\begin{cases}-1, & pm\le e^{3.5}\\ 0, & e^{3.5}<pm\le e^{5}\\ 1, & pm>e^{5}\end{cases} \tag{2}$$

$PM_{2.5}$单值预报模型能较准确体现$PM_{2.5}$浓度的变化趋势，且达到了预期拟合效果，可开展独立预报，但预报准确度有限。同时模型外推是有风险的。当温差变量预报值为负值且对应$PM_{2.5}$浓度预报值过高时需谨慎接受预报结果。

(2) $PM_{2.5}$浓度区间预报模型，简称$PM_{2.5}$区间预报模型，包含精度预报区间$I_1$与准确度预报区间$I_2$。模型结构为：

$$I=\begin{cases}(0,\ 35), & \widehat{pm}<35\\ \left[\widehat{pm}-r,\widehat{pm}+1.5r\right], & 35<=\widehat{pm}<150, \quad r\ge 0\\ >150, & \widehat{pm}>150\end{cases},$$

$$I_1=\begin{cases}(0,\ 35), & \widehat{pm}<35\\ \left[\widehat{pm}-20,\widehat{pm}+30\right], & 35<=\widehat{pm}<150\\ >150, & \widehat{pm}>150\end{cases},$$

$$I_2=\begin{cases}(0,\ 35), & \widehat{pm}<35\\ \left[\widehat{pm}-30,\widehat{pm}+45\right], & 35<=\widehat{pm}<150\\ >150, & \widehat{pm}>150\end{cases} \tag{3}$$

精度预报区间的提议标准为 60%，准确度预报区间的提议标准为 80%。两类预报区间的样本包含率分别为 65.4%，83%，均通过了模型检验。在预报应用中，两类预报区间的样本包含率分别为 52.0%，74.4%，与提议标准有差距，但仍能较好开展预报工作。

(3) NCEP 预报模型，针对 NCEP CFS2 气象预报系统的 $PM_{2.5}$ 浓度预报模型。NCEP 单值预报模型与 $PM_{2.5}$ 单值预报模型相同。NCEP 区间预报模型包含精度预报区间 $I_1$ 与准确度预报区间 $I_2$。模型结构为：

$$\widehat{pm}=\exp\begin{pmatrix}4.567223\exp\left(-0.34431/trg\right)-0.002258w-0.000109t\\ -0.000912pc-0.005976ep+0.736975id\end{pmatrix},$$

$$id=\begin{cases}-1, & pm\le e^{3.5}\\ 0, & e^{3.5}<pm\le e^{5}\\ 1, & pm>e^{5}\end{cases},$$

$$I=\begin{cases}(0,\ 35), & \widehat{pm}<35\\ \left[\widehat{pm}-1.5r,\widehat{pm}+r\right], & 35<=\widehat{pm}<150, \quad r\ge 0\\ >150, & \widehat{pm}>150\end{cases},$$

$$I_1=\begin{cases}(0,\ 35), & \widehat{pm}<35\\ \left[\widehat{pm}-30,\widehat{pm}+20\right], & 35<=\widehat{pm}<150, \quad r\ge 0\\ >150, & \widehat{pm}>150\end{cases},$$

$$I_2 = \begin{cases} (0,\ 35), & \widehat{pm} < 35 \\ \left[\widehat{pm} - 45, \widehat{pm} + 30\right], & 35 <= \widehat{pm} < 150, \quad r \geq 0 \\ > 150, & \widehat{pm} > 150 \end{cases} \tag{4}$$

两类预报区间的样本包含率非常接近提议标准，预报效果显著优于$PM_{2.5}$区间预报模型，可实现独立预报。

在结合 NCEP CFS2 气象预报系统时，建议使用 NCEP 预报模型。在使用 NCEP 预报模型时，建议首先使用 NCEP 区间预报模型，可参考 NCEP 单值预报模型的预报结果。

(4) *id* 变量预报值的两种算法。使用前一天$PM_{2.5}$观测值计算*id*变量预报值的算法便捷快速，但依赖真实数据；使用剔除*id*变量后的$PM_{2.5}$单值预报模型计算*id*变量预报值的算法准确度受限，但能实现独立预报。

可直接采用第一种算法。当$PM_{2.5}$单值预报模型得到进一步优化后，建议采用第二种算法。同时建议根据实际需求选择合适的算法。

# 参考文献

# 致　　谢

感谢陈玉蓉导师给予我充分的支持与信任，在本课题的研究中陈老师提出了许多中肯的建议，并及时指导并配合我开展研究。同时感谢学校、学院的培养与任课老师的教导。特别致谢中国气象数据网与 RDA 的数据支持。

# 附 录 A

## 附 A1 部分建模数据集

表 A1 2014 年 1 月 $PM_{2.5}$ 浓度与各气象观测数据表

| 日 | $PM_{2.5}$浓度 | 平均气温 | 最高气温 | 最低气温 | 20 至 20 时降水量 | 最大风速 | 大型蒸发量 |
|---|---|---|---|---|---|---|---|
| 1 | 153 | 44 | 179 | -26 | 0 | 27 | 17 |
| 2 | 181 | 44 | 155 | -25 | 0 | 21 | 14 |
| 3 | 174 | 60 | 162 | -10 | 0 | 55 | 21 |
| 4 | 112 | 45 | 160 | -11 | 0 | 17 | 15 |
| 5 | 147 | 41 | 161 | -35 | 0 | 40 | 19 |
| 6 | 185 | 69 | 102 | 29 | 10 | 41 | 12 |
| 7 | 268 | 54 | 62 | 50 | 167 | 44 | 22 |
| 8 | 202 | 47 | 78 | 12 | 13 | 57 | 16 |
| 9 | 208 | 17 | 88 | -23 | 0 | 27 | 12 |
| 10 | 128 | 32 | 68 | -12 | 8 | 38 | 13 |
| 11 | 91 | 41 | 52 | 28 | 50 | 30 | 15 |
| 12 | 168 | 48 | 106 | -11 | 0 | 48 | 16 |
| 13 | 266 | 23 | 101 | -23 | 0 | 30 | 13 |
| 14 | 227 | 16 | 109 | -35 | 0 | 32 | 11 |
| 15 | 165 | 21 | 106 | -25 | 0 | 38 | 13 |
| 16 | 122 | 53 | 131 | -4 | 0 | 22 | 15 |
| 17 | 237 | 50 | 135 | -6 | 0 | 26 | 10 |
| 18 | 234 | 23 | 115 | -40 | 0 | 32 | 15 |
| 19 | 221 | 28 | 128 | -27 | 0 | 27 | 20 |
| 20 | 148 | 48 | 146 | -21 | 0 | 49 | 22 |
| 21 | 67 | 11 | 117 | -49 | 0 | 41 | 18 |
| 22 | 91 | 40 | 131 | -47 | 0 | 38 | 15 |
| 23 | 100 | 81 | 164 | -4 | 0 | 45 | 27 |
| 24 | 126 | 111 | 218 | 52 | 0 | 51 | 24 |
| 25 | 248 | 84 | 142 | 31 | 0 | 47 | 21 |
| 26 | 287 | 67 | 138 | 15 | 0 | 27 | 13 |
| 27 | 214 | 66 | 140 | 16 | 0 | 27 | 10 |
| 28 | 284 | 82 | 97 | 42 | 5 | 28 | 3 |
| 29 | 286 | 74 | 98 | 64 | 118 | 38 | 1 |
| 30 | 151 | 85 | 157 | 25 | 2 | 30 | 0 |
| 31 | 177 | 151 | 254 | 73 | 0 | 44 | 0 |
| 单位 | μg/$m^3$ | 0.1℃ | 0.1℃ | 0.1℃ | 0.1mm | 0.1m/s | 0.1mm |

数据来源：中国空气质量在线监测平台、中国气象数据网

## 附 A2 部分预报应用数据集

**表 A2 2017 年 12 月 $PM_{2.5}$ 浓度观测数据与各气象预报数据表**

| 日 | $PM_{2.5}$浓度 | 平均气温 | 最高气温 | 最低气温 | 20 至 20 时降水量 | 最大风速 | 大型蒸发量 |
|---|---|---|---|---|---|---|---|
| 1 | 54 | 73.075 | 112.5 | 83.5 | 0.025 | 24.287 | 11 |
| 2 | 69 | 100.055 | 59.5 | 52.5 | 0.025 | 18.381 | 8 |
| 3 | 84 | 97.210 | 33.5 | 33.5 | 0 | 19.593 | 7 |
| 4 | 180 | 75.593 | 139.5 | 50.5 | 0 | 46.179 | 14 |
| 5 | 162 | 58.695 | 138.5 | 89.5 | 0 | 27.523 | 14 |
| 6 | 137 | 77.448 | 83.5 | 68.5 | 0 | 18.627 | 9 |
| 7 | 95 | 78.368 | 48.5 | 54.5 | 0.025 | 44.110 | 9 |
| 8 | 86 | 45.310 | 136.5 | 64.5 | 0 | 32.275 | 20 |
| 9 | 70 | 57.568 | 140.5 | 91.5 | 0 | 21.645 | 19 |
| 10 | 75 | 60.923 | 72.5 | 85.5 | 0 | 30.513 | 17 |
| 11 | 53 | 68.405 | 71.5 | 66.5 | 0 | 30.206 | 17 |
| 12 | 67 | 69.568 | 148.5 | 82.5 | 0 | 30.825 | 14 |
| 13 | 71 | 48.108 | 143.5 | 105.5 | 2.275 | 43.633 | 9 |
| 14 | 43 | 52.308 | 91.5 | 78.5 | 0.675 | 30.239 | 2 |
| 15 | 56 | 40.060 | 60.5 | 60.5 | 0.150 | 42.954 | 3 |
| 16 | 90 | 29.965 | 147.5 | 75.5 | 0 | 49.343 | 19 |
| 17 | 71 | 19.305 | 155.5 | 108.5 | 0 | 20.537 | 14 |
| 18 | 81 | 27.500 | 87.5 | 78.5 | 0 | 24.945 | 10 |
| 19 | 91 | 37.700 | 69.5 | 58.5 | 0 | 16.832 | 7 |
| 20 | 85 | 54.598 | 160.5 | 74.5 | 0 | 26.737 | 10 |
| 21 | 103 | 68.648 | 163.5 | 94.5 | 0 | 26.565 | 12 |
| 22 | 84 | 93.020 | 69.5 | 54.5 | 0 | 29.130 | 16 |
| 23 | 125 | 86.245 | 28.5 | 24.5 | 0 | 26.848 | 6 |
| 24 | 133 | 69.595 | 105.5 | 36.5 | 0 | 28.485 | 18 |
| 25 | 56 | 62.388 | 114.5 | 73.5 | 0 | 25.389 | 17 |
| 26 | 85 | 76.848 | 54.5 | 52.5 | 0 | 19.443 | 14 |
| 27 | 109 | 87.700 | 42.5 | 35.5 | 0.250 | 27.098 | 9 |
| 28 | 125 | 77.833 | 144.5 | 51.5 | 6.650 | 20.427 | 4 |
| 29 | 144 | 91.293 | 149.5 | 80.5 | 0 | 22.401 | 5 |
| 30 | 143 | 64.803 | 77.5 | 56.5 | 0 | 37.523 | 13 |
| 31 | 130 | 65.753 | 53.5 | 41.5 | 0 | 11.729 | 17 |
| 单位 | $\mu g/m^3$ | 0.1℃ | 0.1℃ | 0.1℃ | 0.1mm | 0.1m/s | 0.1mm |

数据来源：中国空气质量在线监测平台、中国气象数据网、RDA

# 附 录 B

附注根据文献[12]编写的计算 Bates 曲率度量与 Box 偏差度量的 R 函数，以及如何使用该函数的 R 程序说明。

## 附 B1 Bates 曲率度量与 Box 偏差度量程序

函数 noncurva(nl.1,V1,V2,alpha=0.05)用于计算 Bates 曲率与 Box 偏差，适用于未知参数个数大于等于 2 的非线性回归模型。其中，nl.1 表示非线性回归模型,V1 表示一阶导矩阵 $\dot{\mathbf{V}}$ ,V2 表示二阶导立体阵 $\ddot{\mathbf{V}}$ 。程序如下：

```
noncurva=function(nl.1,V1,V2,alpha=0.05){
   qr=qr(V1)
   Q=qr.Q(qr,complete = T)
   R1=qr.R(qr)
   L=solve(R1)

   U=V2
   for(i in 1:dim(V2)[1])    U[i,,]=t(L)%*%V2[i,,]%*%L

   A=U
   for(i in 1:dim(U)[3])    A[,,i]=t(Q)%*%U[,,i]

   p=dim(V1)[2]
   A1=A[1:p,,]
   A2=A[-c(1:p),,]

   coef=coef(nl.1)
   sigma=sigma(nl.1)
   rho=sigma*sqrt(length(coef))

   #mean square curvature
   rms=function(At){
      rms0=rep(0,dim(At)[1])
      for(i in 1:dim(At)[1])    rms0[i]=2*sum(At[i,,]^2)+(sum(diag(At[i,,])))^2
      rms2=sum(rms0)/(p*(p+2))
      rms=sqrt(rms2)*rho
      return(round(rms,5))
   }
   rms.t=rms(A1)
   rms.n=rms(A2)
```

```
  #standard
  n=dim(A)[1]
  rf=qf(1-alpha,p,n-p)
  rf0=1/sqrt(rf)
  rf1=0.5/sqrt(rf)
  rf2=0.2/sqrt(rf)
  bound.25.10=c(round(rf0,5),round(rf1,5),round(rf2,5))

  #box bias
  w=rep(0,p)
  for(i in 1:n){
    v=V1[i,]*sum(diag(U[i,,]))
    w=w+v
  }
  bias.coef=-L%*%t(L)%*%w*sigma^2/2
  bias.coef2=round(bias.coef,5)
  percentbias.coef=bias.coef/coef*100
  percentbias.coef2=round(percentbias.coef,5)

  return(list(KMS.P=rms.t,KMS.N=rms.n,Bound=bound.25.10,
              Bias.coef=bias.coef2,PercentBias.coef=percentbias.coef2))
}
```

## 附 B2 非线性回归建模程序

此部分说明如何根据非线性回归模型使用 noncurva 函数。程序如下：

```
nl.1=nls(lpm~a*exp(-b/trg)+c*w+d*t+e*pc+f*ep+g*id,
         start = list(a=40,b=1,c=0,d=0,e=0,f=0,g=1),trace = T) #nonlinear model

forl=formula(nl.1)
coef=coef(nl.1)
xlist=c("a","b","c","d","e","f","g")

fx=deriv(forl,xlist,hessian = T,function.arg = T )
V1=attr(fx(coef[1],coef[2],coef[3],coef[4],coef[5],coef[6],coef[7]),"gradient")
V2=attr(fx(coef[1],coef[2],coef[3],coef[4],coef[5],coef[6],coef[7]),"hessian")
noncurva(nl.1,V1,V2)
```